\documentclass[aps,prd,preprint,groupedaddress]{revtex4-1}
\usepackage{extraipa}
\usepackage{amsmath}
\usepackage{amsfonts}
\usepackage{amssymb}
\usepackage{bbold}

\newcommand{\intp}[1]{\int\frac{d^4{#1}}{(2\pi)^4}}
\newcommand{\intx}[1]{\int d^4{#1} }
\newcommand{\etal}{\textit{et al}.}

\begin{document}

\title{Covariant quantum corrections to a scalar field model inspired by nonminimal natural inflation}

\author{Sandeep Aashish}%
\email[]{sandeepa16@iiserb.ac.in}

\author{Sukanta Panda}
\email[]{sukanta@iiserb.ac.in}

\affiliation{Department of Physics, Indian Institute of Science Education and Research, Bhopal 462066, India}

%


\date{\today}

\begin{abstract}
We calculate the covariant one-loop quantum gravitational effective action for a scalar field model inspired by the recently proposed nonminimal natural inflation model. Our calculation is perturbative, in the sense that the effective action is evaluated in orders of background field, around a Minkowski background. The effective potential has been evaluated taking into account the finite corrections. An order-of-magnitude estimate of the one-loop corrections reveals that gravitational and non-gravitational corrections have same or comparable magnitudes. 
\end{abstract}



\maketitle

\section{\label{intro}Introduction}
A fully consistent quantum theory of gravity has remained elusive despite longstanding efforts to construct a gravity theory that is valid at the Planck scale (see Ref. \cite{woodard2009} for a comprehensive review). A well known problem with quantizing gravity perturbatively is the inability to consistently absorb the divergences, giving rise to non-renormalizability. However, in the past two decades or so, a more modern view has developed where general relativity is studied as a quantum effective field theory at low energies \cite{donoghue1994}. This treatment allows separation of quantum effects from known low energy physics from those that depend on the ultimate high energy completion of the theory of gravity (see Ref. \cite{donoghue2015} for a review by Donoghue and Holstein). 

More recently, with the availability of high precision data from experiments probing the early universe, especially inflation era, it has become important to consider quantum gravitational corrections in early universe cosmology \cite{fabris2012,krauss2014,woodard2014}. This has motivated several studies of aspects of quantum gravitational corrections in inflationary universe, see for example Refs. \cite{klemm2004,cognola2005,cognola2012,hebecker2017,herranen2017,bounakis2018,markkanen2018,ruf2018,heisenberg2019}.

One of the well-known methods employed in such studies is to compute the effective action, which is known to be the generator of 1PI diagrams \cite{buchbinder1992,vilkovisky1992}. An advantage of this technique is that one can directly obtain divergence structure at a given loop order, without going through the hassle of summing over individual Feynman diagram contributions. Other applications include the calculation of effective potential \cite{coleman1973}. The computation of effective action is most commonly carried out using the background field method, where small fluctuations about a classical background field are quantized, not the total field. In general, it turns out that the results consequently depend on the choice of background field \cite{falkenberg1998,labus2016,ohta2016}. In case of gravity, which is treated as a gauge theory, it is therefore important to ensure that there are no fictitious dependence of conclusions on the choice of gauge and background. Hence, in this work, we use DeWitt-Vilkovisky's covariant effective action approach that systematically yields gauge and background independent effective action \cite{dewitt1967b,*dewitt1967c,parker2009}.  

We consider a recently proposed modification of the natural inflation (NI) model \cite{freese1990}, wherein a periodic nonminimal coupling term similar to NI potential is added along with a new parameter, that eventually leads to a better fit with Planck results \cite{ferreira2018}. These phenomenological implications are in no way the only motivation for considering this model in the present work. Rather, it serves as a toy model to achieve our mainly three objectives, which are as follows. First, to set up the computation using symbolic manipulation packages to evaluate one-loop covariant effective action up to quartic order terms in the background field. As a starting point, we work in the Minkowski background. Second, we aim to recover and establish past results. And third, we wish to estimate the magnitudes of quantum gravitational corrections from the finite contributions at least for the effective potential, since there are typically several thousands of terms one has to deal with.

The organization of this paper is as follows. In Sec. \ref{sec2}, we introduce and briefly review the nonminimal natural inflation model. Sec. \ref{sec3} covers a review of covariant effective action formalism, notations, and the methodology of our calculations. Sec. \ref{sec4} constitutes a major part of this paper, detailing the calculations of each contributing term mentioned in Sec. \ref{sec3}, along with the divergent part, loop integrals, and renormalization. Some past results and their extensions have also been presented. Finally, in Sec. \ref{sec5}, we derive the effective potential including the finite corrections from the loop integrals, and perform an order-of-magnitude estimation of quantum corrections.

\section{\label{sec2}Periodic nonminimal natural inflation model}
Natural inflation was first introduced by Freese \etal \cite{freese1990} as an approach where inflation arises dynamically (or \textit{naturally}) from particle physics models. In natural inflation models, a flat potential is effected using pseudo Nambu-Goldstone bosons arising from breaking the continuous shift symmetry of Nambu-Goldstone modes into a discrete shift symmetry. As a result, the inflation potential in a Natural inflation model takes the form,
\begin{eqnarray}
\label{eq01}
V(\phi) = \Lambda^4 \left(1 + \cos(\phi/f)\right);
\end{eqnarray} 
where the magnitude of parameter $\Lambda^4$ and periodicity scale $f$ are model dependent. However, majority of natural inflation models are in tension with recent Planck 2018 results \cite{planck2018x}. However, it was shown in Ref. \cite{gerbino2017} that once neutrino properties are more consistently taken into account when analyzing the data, natural inflation does marginally agree with data.

This work concerns a recently proposed extension of the original natural inflation model introducing a new periodic non-minimal coupling to gravity \cite{ferreira2018}. The authors in \cite{ferreira2018} showed that the new model leads to a better fit with observation data thanks to the introduction of a new parameter in the nonminimal coupling term, with $n_{s}$ and $r$ values well within $95\%$ C.L. region from combined Planck 2018+BAO+BK14 data. An important feature of this model is that $f$ becomes sub-Planckian, contrary to a super-Planckian $f$ in the original natural inflation model \cite{freese1990}, and thus addresses issues related to gravitational instanton corrections \cite{banks2003,rudelius2015a,rudelius2015b,montero2015,hebecker2017}.
 
Our objective here is to study one-loop quantum gravitational corrections to the natural inflation model with non-minimal coupling, using Vilkovisky-DeWitt's covariant effective action approach \cite{parker2009}. One of the first works considering one-loop gravitational corrections were pioneered by Elizalde and Odintsov \cite{elizalde1993,elizalde1994b,elizalde1994c,elizalde1994d}. Vilkovisky-DeWitt method was used to study effective actions in Refs. \cite{odintsov1989,odintsov1990a,odintsov1990b,odintsov1991,odintsov1993}. Unfortunately, calculating the covariant effective action exactly is highly nontrivial, though non-covariant effective actions can in principle be evaluated using proper time methods. Hence, we take a different route by employing a perturbative calculation of one-loop effective action, in orders of the background scalar field. This requires us to apply a couple of approximations. First, we work in the regime where potential is flat, i.e. $\phi \ll f$, which is generally true during slow-rolling inflation. Second, the background metric is set to be Minkowski. This choice is debatable, since it does not accurately represent an inflationary scenario, but has been used before \cite{saltas2017,bounakis2018} as a first step towards studying quantum corrections.

The action for the nonminimal natural inflation in the Einstein frame is given by,
\begin{eqnarray}
\label{eq02}
S = \int d^4 x \sqrt{-g}\left(-\dfrac{2 R}{\kappa^2} + \dfrac{1}{2}K(\phi)\phi {}_{;a} \phi {}^{;a} + \dfrac{V(\phi)}{(\gamma(\phi))^4}\right)
\end{eqnarray}
where, 
\begin{equation}
\label{eq03}
\gamma(\phi)^2 = 1 + \alpha\left(1+\cos\left(\dfrac{\phi}{f}\right)\right),
\end{equation}
and,
\begin{equation}
\label{eq04}
K(\phi) = \dfrac{1 + 24\gamma'(\phi)^2/\kappa^2}{\gamma(\phi)^2}. 
\end{equation}
$V(\phi)$ is as in Eq. (\ref{eq01}). Here, $\phi_{;a} \equiv \nabla_{a}\phi$ denotes the covariant derivative.
In the region where potential is flat, $\phi/f \ll 1$, and we expand all periodic functions in Eq. (\ref{eq02}) up to quartic order in $\phi$ followed by rescaling $\sqrt{k_0}\phi \to \phi$:
\begin{equation}
\label{action}
S \approx \int d^4 x \sqrt{-g} \left(- \frac{2 R}{\kappa^2} + \tfrac{1}{2} \frac{m^2}{k_0} \phi^2 + \tfrac{1}{24} \frac{\lambda}{k_0^2} \phi^4 + \tfrac{1}{2} (1 + \frac{k_1}{k_0^2} \phi^2) \phi {}_{;a} \phi {}^{;a}\right) + \mathcal{O}(\phi^5)
\end{equation}
where parameters $m, \lambda,k_0$ and $k_1$ have been defined out of $\alpha, f$ and $\Lambda^4$ in from Eq. (\ref{eq02}):
\begin{eqnarray}
\label{param}
m^2 &=& \dfrac{\Lambda^4 (2\alpha - 1)}{(1 + 2\alpha)^3 f^2};\nonumber \\
\lambda &=& \dfrac{\Lambda^4 (8\alpha^2 - 12\alpha + 1)}{(1 + 2\alpha)^4 f^4}; \nonumber \\
k_0 &=& \dfrac{1}{1 + 2\alpha}; \nonumber \\
k_1 &=& \dfrac{\alpha(\kappa^2 f^2 + 96\alpha^2 + 48\alpha)}{2\kappa^2 f^4 (1 + 2\alpha)^2}. \nonumber \\
\end{eqnarray}
We have also omitted a constant term appearing in ($\ref{action}$) because such terms are negligibly small in early universe. The action (\ref{action}) is in effect a $\phi^4$ scalar theory with derivative coupling.

\section{\label{sec3}Effective action formalism}
A standard procedure while calculating loop corrections in quantum field theory, is to use the well known background field method, according to which a field is split into a classical background and a quantum part that is much smaller in magnitude (and hence treated perturbatively) \cite{falkenberg1998,labus2016,ohta2016}. A by-product of this procedure is the background and gauge dependence of quantum corrections. We briefly review here the covariant effective action formalism, that yields gauge-invariant and background field independent results, employed in this work. Interested reader is advised to see Ref. \cite{parker2009} (chap. 7) for a detailed review. 

Quantization of a theory $S[\varphi]$ with fields $\varphi^{i}$ is performed about a classical background $\bar{\varphi}^{i}$:  $\varphi^{i} = \bar{\varphi}^{i} + \zeta^{i}$, where $\zeta^{i}$ is the quantum part. Here, $\varphi^i$ is the local coordinate of a point in the `field space' and represents any scalar or vector or tensor field(s) in the coordinate space. The index $i$ in field space corresponds to all gauge indices and coordinate dependence of fields. This way of writing the field-space equivalent of a coordinate space quantity (such as a vector field) is called condensed notation \cite{dewitt1964}. In our case, $\varphi^{i}=\{g_{\mu\nu}(x),\phi(x)\}$; $\bar{\varphi}^{i}=\{\eta_{\mu\nu},\bar{\phi}(x)\}$ where $\eta_{\mu\nu}$ is the Minkowski metric; and, $\zeta^{i}=\{\kappa h_{\mu\nu}(x),\delta\phi(x)\}$. The fluctuations $\zeta^{i}$ are assumed to be small enough for a perturbative treatment to be valid, viz. $|\kappa h_{\mu\nu}|\ll 1; |\delta\phi|\ll |\phi|$. In this limit, the infinitesimal general coordinate transformations can be treated as gauge transformations associated with $h_{\mu\nu}$ \cite{donoghue1994,donoghue2017}. In fact, for any metric $g_{\mu\nu}(x)$, this infinitesimal transformation takes the form,
\begin{eqnarray}
    \label{aeaf0}
    \delta g_{\mu\nu} = -\delta\epsilon^\lambda g_{\mu\nu,\lambda}-\delta\epsilon^\lambda \,_{,\mu}g_{\lambda\nu}-\delta\epsilon^\lambda \,_{,\nu}g_{\lambda\mu}.
\end{eqnarray}
In the condensed notation, an infinitesimal gauge transformation of any field $\varphi^i$ is given by,
\begin{eqnarray}
    \label{aeaf1}
    \delta\varphi^{i} = K^{i}_{\alpha}[\varphi]\delta\epsilon^{\alpha},
\end{eqnarray}
where $K^{i}_{\alpha}$ is identified as the generator of gauge transformations, while $\delta\epsilon^{\alpha}$ are the gauge parameters. As with Einstein notations, repeated (or contracted) indices in the condensed notation represent a sum over all the associated gauge or tensor indices and integral over all coordinate indices. The gauge fixing condition is given by fixing a functional $\chi_{\alpha}[\bar{\varphi}]$ so that it intersects each gauge orbit in field space only once. Including the gauge-fixing condition(s) and corresponding ghost determinant(s), the covariant one-loop effective action is given by \cite{huggins1987,toms2007}
\begin{eqnarray}
\label{aeaf2}
\Gamma = -\ln\int[d\zeta]\exp\left[\dfrac{1}{2}\left(-\zeta^{i}\zeta^{j}\Big(S_{,ij}[\bar{\varphi}] - \Gamma^{k}_{ij}S_{,k}[\bar{\varphi}]\Big) - \frac{1}{2\alpha}f_{\alpha\beta}\chi^{\alpha}\chi^{\beta}\right)\right]-\ln\det Q_{\alpha\beta}[\bar{\varphi}],
\end{eqnarray}
as $\alpha\longrightarrow 0$ (Landau gauge). Here, $[d\zeta]\equiv \prod_{i} d\zeta$. A few comments on Eq. (\ref{aeaf2}) are in order. The first term inside the exponential is the covariant derivative of the action functional with respect to $\zeta^{i}$ in field space. $\Gamma^{k}_{ij}$ are the field-space connections defined with respect to the field-space metric $G_{ij}$, and are responsible for general covariance of Eq. (\ref{aeaf2}). In general, the field-space connections have complicated, non-local structure especially in presence of a gauge symmetry. However, they reduce to the standard Christoffel connections, in terms of $G_{ij}$, when $\chi_{\alpha}$ is chosen to be the Landau-DeWitt gauge i.e. $\chi_{\alpha} = K_{\alpha i}[\bar{\varphi}]\zeta^{i} = 0$, along with $\alpha\to 0$ \cite{mackay2010,bounakis2018}. $f_{\alpha\beta}$ is any symmetric, positive definite operator and makes no non-trivial contribution to effective action \cite{parker2009}. Note also that the contributions from connection terms, and hence the question of covariance, is relevant for off-shell analyses, since $S_{,i} = 0$ on-shell. $\det Q_{\alpha\beta}$ is the ghost determinant term that appears during quantization. This term is absorbed into the exponential by introducing Faddeev-Popov ghosts, $c^{\alpha}$ and $\bar{c}^{\alpha}$, so that \cite{parker2009},
\begin{eqnarray}
    \label{aeaf3}
    \ln\det Q_{\alpha\beta} = \ln\int [d \bar{c}^{\alpha}] [d c^{\beta}] \exp \left(- \bar{c}^{\alpha} Q_{\alpha\beta} c^{\beta} \right).
\end{eqnarray}
As a result, 
\begin{eqnarray}
    \label{aeaf4}
    \Gamma[\bar{\varphi}] = -\ln\int [d\zeta] [d \bar{c}^{\alpha}] [d c^{\beta}] \exp\left[-\dfrac{\zeta^{i}\zeta^{j}}{2}\Big(S_{,ij}[\bar{\varphi}] - \Gamma^{k}_{ij}S_{,k}[\bar{\varphi}]\Big) - \frac{1}{4\alpha}f_{\alpha\beta}\chi^{\alpha}\chi^{\beta} - \bar{c}^{\alpha} Q_{\alpha\beta} c^{\beta}\right].
\end{eqnarray}

The computation of Eq. (\ref{aeaf4}) traditionally has involved the use of proper time method, such as employing the heat kernel technique. For Laplace type operators (coefficients of $\zeta^{i}\zeta^{j}$ in the exponential), of the form $g^{\mu\nu}\nabla_{\mu}\nabla_{\nu} + Q$ (where $Q$ does not contain derivatives), the heat kernel coefficients are known and are quite useful because they are independent of dimensionality \cite{dewitt1964}. However, these operators in general are not Laplace type, as in the present case. A class of nonminimal operators such as the one in Eq. (\ref{aeaf4}) can be transformed to minimal (Laplace) form using the generalised Schwinger-DeWitt technique \cite{barvinsky1985}, but in practice the implementation is quite complicated and specific to a given Lagrangian. Examples of such an implementation can be found in Refs. \cite{alvarez2015,steinwachs2011}. We take a different approach here, calculating the one-loop effective action perturbatively in orders of the background field. While one does not obtain exact results in a perturbative approach, unlike the heat kernel approach, it is possible to obtain accurate results up to a certain order in background fields which is of relevance for a theory in, say, the early universe. Some past examples are Refs. \cite{saltas2017,bounakis2018}. Moreover, our implementation of this method using xAct packages for Mathematica \cite{xact,xpert} is fairly general in terms of its applicability to not only scalars coupled with gravity, but also vector and tensor fields (see, for instance, Ref. \cite{aashish2019b}). A caveat at this time, is that the perturbative expansions are performed about the Minkowski background and not a general metric background. However, a generalization to include FRW background is part our future plans. 

For convenience, we write the exponential in the first term of $\Gamma$ as,
\begin{eqnarray}
\label{aeaf5}
\exp[\cdots] &=& \exp\left\{- \left(\tilde{S}[\bar{\varphi}^{0}]+\tilde{S}[\bar{\varphi}^{1}]+\tilde{S}[\bar{\varphi}^{2}]+\tilde{S}[\bar{\varphi}^{3}]+\tilde{S}[\bar{\varphi}^{4}]\right)\right\}\nonumber \\
&\equiv & \exp\left\{- (\tilde{S}_{0}+\tilde{S}_{1}+\tilde{S}_{2}+\tilde{S}_{3}+\tilde{S}_{4})\right\}
\end{eqnarray}
$\tilde{S}_0$ yields the propagator for each of the fields $\zeta^{i}$. The rest of the terms are contributions from interaction terms, which we assume to be small. Treating $\tilde{S}_{1},...,\tilde{S}_{4}$ as perturbative, and expanding Eq. (\ref{aeaf5}), $\Gamma[\bar{\varphi}]$ can be written as,
\begin{eqnarray}
    \label{aeaf6}
    \Gamma[\bar{\varphi}] &=& -\ln\int [d\zeta] [d \bar{c}^{\alpha}] [d c^{\beta}] e^{-S_0}(1-\delta S + \dfrac{\delta S^2}{2} + \cdots) ; \nonumber \\
    &=& - \ln (1 - \langle\delta S\rangle + \dfrac{1}{2}\langle\delta S^2 \rangle + \cdots); 
\end{eqnarray}
where $\delta S = \sum_{i=1}^4 \tilde{S}_{i}$ and $\langle\cdot\rangle$ represents the expectation value in the path integral formulation. Finally, we use $\ln(1 + x)\approx x$ to find the contributions to $\Gamma$ at each order of background field. We only use the leading term in the logarithmic expansion, since all higher order terms will yield contributions from disconnected diagrams viz-a-viz $\langle\delta S\rangle^{2},$ etc which we ignore throughout our calculation. Moreover, since we are interested in terms up to quartic order in background field, we truncate the Taylor series in Eq. (\ref{aeaf6}) up to $\delta S^{4}$. With these considerations, the final contributions to $\Gamma$ at each order of $\bar{\varphi}$ is:
\begin{eqnarray}
\label{aeaf7}
\mathcal{O}(\bar{\varphi}) &:& \langle\tilde{S}_{1}\rangle;\nonumber \\
\mathcal{O}(\bar{\varphi}^2) &:& \langle\tilde{S}_{2}\rangle - \dfrac{1}{2}\langle\tilde{S}_{1}^{2}\rangle; \nonumber \\
\mathcal{O}(\bar{\varphi}^3) &:& \langle\tilde{S}_{3}\rangle - \langle\tilde{S}_{1}\tilde{S}_{2}\rangle + \dfrac{1}{6} \langle\tilde{S}_{1}^{3}\rangle;\nonumber \\
\mathcal{O}(\bar{\varphi}^4) &:& \langle\tilde{S}_{4}\rangle - \langle\tilde{S}_{1}\tilde{S}_{3}\rangle + \dfrac{1}{2}\langle\tilde{S}_{1}^{2}\tilde{S}_{2}\rangle -\dfrac{1}{2}\langle\tilde{S}_{2}^{2}\rangle - \dfrac{1}{24}\langle\tilde{S}_{1}^{4}\rangle .
\end{eqnarray}
Also, we recall that the metric fluctuations have a factor of $\kappa$. Accordingly, the terms in Eq. (\ref{aeaf7}) will also contain powers of $\kappa$. It turns out, as will be shown below, that all contributions are at most of the order $\kappa^4$. Expecting $\mathcal{O}(\kappa^4)$ terms to be significantly suppressed, we only take into account the corrections up to $\mathcal{O}(\kappa^2)$. In what follows, we will detail the evaluation of terms in Eq. (\ref{aeaf7}).


\section{\label{sec4}Covariant one-loop corrections}
\subsection{Setup}
The first step towards writing $\Gamma[\bar{\varphi}]$ in Eq. (\ref{aeaf4}) is to identify the field space metric, given in terms of the field-space line element, 
\begin{align}
    ds^{2} &= G_{ij}d\varphi^{i} d\varphi^{j} \\ 
    \label{acqc0}
    &= \int d^{4}x d^{4}x' \left(G_{g_{\mu\nu}(x)g_{\rho\sigma}(x')}dg_{\mu\nu}(x) dg_{\rho\sigma}(x') + G_{\phi(x)\phi(x')} d\phi(x) d\phi(x')\right).
\end{align}
A prescription for identifying field space metric is to read off the components of $G_{ij}$ from the coefficients of highest derivative terms in classical action functional \cite{vilkovisky1984b}. For the scalar field $\phi(x)$, the field-space metric is chosen to be,
\begin{eqnarray}
    \label{acqc1}
    G_{\phi(x)\phi(x')}=\sqrt{g(x)}\delta(x,x');
\end{eqnarray}
For the metric $g_{\mu\nu}(x)$, a standard choice for field-space metric is \cite{parker2009,mackay2010}
\begin{align}
    \label{acqc2}
    G_{g_{\mu\nu}(x)g_{\rho\sigma}(x')}=\dfrac{\sqrt{g(x)}}{\kappa^2}\left\{ g^{\mu(\rho}(x)g^{\sigma)\nu}(x)- \frac{1}{2}g^{\mu\nu}(x)g^{\rho\sigma}(x) \right\}\delta(x,x'),
\end{align}
where the brackets around tensor indices in the first term indicate symmetrization. As a convention, we choose to include $\kappa^2$ factor in Eq. (\ref{acqc2}) to account for dimensionality of the length element in Eq. (\ref{acqc0}), although choosing otherwise is also equally valid as long as dimensionality is taken care of. The inverse metric can be derived from the identity $G_{ij}G^{jk}=\delta^{k}_{i}$:
\begin{eqnarray}
    \label{acqc3}
    G^{g_{\mu\nu}(x)g_{\rho\sigma}(x')}&=&\kappa^2\left\{ g_{\mu(\rho}(x)g_{\sigma)\nu}(x)- \frac{1}{2}g_{\mu\nu}(x)g_{\rho\sigma}(x) \right\}\delta(x,x'); \\
    \label{acqc4}
    G^{\phi(x)\phi(x')}&=&\delta(x,x').
\end{eqnarray}
Next, using Eqs. (\ref{acqc1})-(\ref{acqc4}), one can find the Vilkovisky-DeWitt connections $\Gamma^{k}_{ij}$ which has an identical definition to the Christoffel connections thanks to the Landau-DeWitt gauge choice. Out of a total of six possibilities there are three non-zero connections obtained as follows:
\begin{align}
    \label{acqc5}
    \Gamma^{g_{\lambda\tau}(x)}_{g_{\mu\nu}(x')g_{\rho\sigma}(x'')} &= \delta(x'',x')\delta(x'',x)\left[-\delta^{(\mu}_{(\lambda}g^{\nu)(\rho}(x)\delta^{\sigma)}_{\tau)} + \dfrac{1}{4}g^{\mu\nu}(x)\delta^{\rho}_{(\lambda}\delta^{\sigma}_{\tau)} + \dfrac{1}{4}g^{\rho\sigma}(x)\delta^{\mu}_{(\lambda}\delta^{\nu}_{\tau)} \right. \nonumber \\ & \left. + \dfrac{1}{4}g_{\lambda\tau}(x)g^{\mu(\rho}(x)g^{\sigma)\nu}(x) - \dfrac{1}{8} g_{\lambda\tau}(x)g^{\mu\nu}(x)g^{\rho\sigma}(x) \right] \\
    \Gamma^{g_{\lambda\tau}(x)}_{\phi(x')\phi(x'')} &= \dfrac{\kappa^2}{4}\delta(x'',x')\delta(x'',x) g_{\lambda\tau}(x) \\
    \Gamma^{\phi(x)}_{\phi(x') g_{\lambda\tau}(x'')} &= \dfrac{1}{4}\delta(x'',x')\delta(x'',x) g^{\lambda\tau}(x) = \Gamma^{\phi(x)}_{g_{\lambda\tau}(x') \phi(x'')}.
\end{align}
Note that upon substituting into Eq. (\ref{aeaf4}), all calculations here are evaluated at the background field(s) which in our case is the Minkowski metric and a scalar field $\bar{\phi}(x)$. We also recall that this rather unrestricted choice of background is allowed because of the DeWitt connections that ensure gauge and background independence. As alluded to earlier, the Landau-DeWitt gauge condition, $K_{\alpha i}[\bar{\varphi}]\zeta^{i} = 0$, is given in terms of the gauge generators $K_{\alpha i}$. Since there is only one set of transformations vi-a-viz general coordinate transformation, there exists one gauge parameter which we call $\xi^{\lambda}(x)$. In the condensed notation, this corresponds to $\delta\epsilon^{\alpha}$ where $\alpha\to (\lambda,x)$. Gauge generator on the gravity side $K^{g_{\mu\nu}}_{\lambda}(x,x')$ is read off from Eq. (\ref{aeaf0}), while $K^{\phi}_{\lambda}(x,x')$ is read off from the transformation of $\phi$:
\begin{eqnarray}
    \label{acqc6}
    \delta_{\xi}\phi = -\partial_{\mu}\phi \xi^{\lambda} .
\end{eqnarray}
Substituting in the definition of $\chi_{\alpha}[\bar{\varphi}]$ in coordinate space, we obtain
\begin{eqnarray}
    \label{acqc7}
    \chi_{\lambda}[\bar{\phi}] &=& \int d^{4}x' \left( K_{g_{\mu\nu} \lambda}(x,x') \kappa h_{\mu\nu}(x') + K_{\phi \lambda}(x,x')\delta\phi(x') \right) \nonumber \\
    &=& \dfrac{2}{\kappa}\left(\partial^{\mu}h_{\mu\lambda} - \dfrac{1}{2}\partial_{\lambda}h\right) - \omega\partial_{\lambda}\bar{\phi}\delta\phi .
\end{eqnarray}
where $\omega$ is a bookkeeping parameter, which we adopt from Ref. \cite{mackay2010}; a second such parameter $\nu$ (not to be confused with the tensor index) appears with all Vilkovisky-DeWitt connection terms to keep track of gauge (non-)invariance. That is, we write $S_{;ij}=S_{,ij} - \nu \Gamma^{k}_{ij}S_{,k}$. As shown later, playing with these parameters reproduces past non-gauge-invariant results. Here and throughout, the indices of field-space quantities like the gauge generator are raised and lowered using field-space metric defined in Eqs. (\ref{acqc1}) - (\ref{acqc4}). Lastly, we choose $f^{\alpha\beta} = \kappa^{2}\eta^{\lambda \lambda'}\delta(x,x')$ in Eq. (\ref{aeaf4}) to determine the gauge fixing term. One last piece needed before background-field-order expansions, the ghost term $Q_{\alpha\beta}$. We use the definition \cite{parker2009}, $Q_{\alpha\beta} \equiv \chi_{\alpha, i}K^{i}_{\beta}$, to obtain
\begin{eqnarray}
    \label{acqc8}
    Q_{\mu\nu} = \left(-\dfrac{2}{\kappa} \eta_{\mu\nu}\partial_{\alpha}\partial^{\alpha} + \omega\partial_{\mu}\bar{\phi}\partial_{\nu}\bar{\phi}\right)\delta(x,x').
\end{eqnarray}

\subsection{\label{loopint}Loop integrals and divergent parts}
Substituting the connections, the gauge fixing term and the ghost term along with action (\ref{action}) in Eq. (\ref{aeaf4}), and employing the notations in Eq. (\ref{aeaf5}), we obtain:
\begin{eqnarray}
    \label{s0}
    \tilde{S}_{0} &=& \intx{x} \Big[\frac{m^2 (\delta \phi)^2}{2 k_0} + \tfrac{1}{2} \delta \phi{}_{,a} \delta \phi{}^{,a} - 2 h^{ab} h^{c}{}_{c}{}_{,a}{}_{,b} -  \frac{2 \bar{c}^{a} c_{a}{}^{,b}{}_{,b}}{\kappa} + h^{ab} h^{}{}_{a}{}^{c}{}_{,b}{}_{,c} -  \frac{h^{ab} h^{}{}_{a}{}^{c}{}_{,b}{}_{,c}}{\alpha} \nonumber \\ 
&& + h^{a}{}_{a} h^{bc}{}_{,b}{}_{,c} + \frac{h^{a}{}_{a} h^{bc}{}_{,b}{}_{,c}}{\alpha}  -  \tfrac{1}{2} h^{ab} h^{}{}_{ab}{}^{,c}{}_{,c} + \tfrac{1}{2} h^{a}{}_{a} h^{b}{}_{b}{}^{,c}{}_{,c} -  \frac{h^{a}{}_{a} h^{b}{}_{b}{}^{,c}{}_{,c}}{4 \alpha} \Big] \\
\label{s1}
\tilde{S}_{1} &=& \intx{x} \Bigl[\frac{m^2 \kappa \delta \phi h^{a}{}_{a} \bar{\phi}}{2 k_0} -  \frac{m^2 \kappa \nu \delta \phi h^{a}{}_{a} \bar{\phi}}{4 k_0} -  \tfrac{1}{2} \kappa \delta \phi h^{b}{}_{b} \bar{\phi} {}^{,a}{}_{,a} + \tfrac{1}{4} \kappa \nu \delta \phi h^{b}{}_{b} \bar{\phi} {}^{,a}{}_{,a} -  \tfrac{1}{2} \kappa \delta \phi h^{b}{}_{b}{}_{,a} \bar{\phi} {}^{,a} \nonumber \\ 
&& + \frac{\kappa \omega \delta \phi h^{b}{}_{b}{}_{,a} \bar{\phi} {}^{,a}}{2 \alpha} + \kappa \delta \phi \bar{\phi} {}^{,a} h^{}{}_{a}{}^{b}{}_{,b} -  \frac{\kappa \omega \delta \phi \bar{\phi} {}^{,a} h^{}{}_{a}{}^{b}{}_{,b}}{\alpha} + \kappa \delta \phi h^{}{}_{ab} \bar{\phi} {}^{,a}{}^{,b}\Bigr] \\
\label{s2}
\tilde{S}_{2} &=& \intx{x} \Bigl[ -\frac{m^2 \kappa^2 h^{}{}_{ab} h^{ab} \bar{\phi}^2}{8 k_0} + \frac{m^2 \kappa^2 h^{a}{}_{a} h^{b}{}_{b} \bar{\phi}^2}{16 k_0} + \frac{\lambda \bar{\phi}^2 (\delta \phi)^2}{4 k_0^2} -  \frac{m^2 \kappa^2 \nu \bar{\phi}^2 (\delta \phi)^2}{8 k_0} \nonumber \\ 
&& + \frac{k_1 \bar{\phi}^2 \delta \phi{}_{,a} \delta \phi{}^{,a}}{2 k_0^2} + \frac{2 k_1 \delta \phi \bar{\phi} \bar{\phi} {}_{,a} \delta \phi{}^{,a}}{k_0^2} -  \tfrac{1}{8} \kappa^2 h^{}{}_{bc} h^{bc} \bar{\phi} {}_{,a} \bar{\phi} {}^{,a} + \tfrac{1}{16} \kappa^2 \nu h^{}{}_{bc} h^{bc} \bar{\phi} {}_{,a} \bar{\phi} {}^{,a} \nonumber \\ 
&& + \tfrac{1}{16} \kappa^2 h^{b}{}_{b} h^{c}{}_{c} \bar{\phi} {}_{,a} \bar{\phi} {}^{,a} -  \tfrac{1}{32} \kappa^2 \nu h^{b}{}_{b} h^{c}{}_{c} \bar{\phi} {}_{,a} \bar{\phi} {}^{,a} + \frac{k_1 (\delta \phi)^2 \bar{\phi} {}_{,a} \bar{\phi} {}^{,a}}{2 k_0^2} -  \tfrac{1}{16} \kappa^2 \nu (\delta \phi)^2 \bar{\phi} {}_{,a} \bar{\phi} {}^{,a} \nonumber \\ 
&& + \frac{\kappa^2 \omega^2 (\delta \phi)^2 \bar{\phi} {}_{,a} \bar{\phi} {}^{,a}}{4 \alpha} + \omega c^{a} \bar{c}^{b} \bar{\phi} {}_{,a} \bar{\phi} {}_{,b} + \tfrac{1}{2} \kappa^2 h^{}{}_{a}{}^{c} h^{}{}_{bc} \bar{\phi} {}^{,a} \bar{\phi} {}^{,b} -  \tfrac{1}{4} \kappa^2 \nu h^{}{}_{a}{}^{c} h^{}{}_{bc} \bar{\phi} {}^{,a} \bar{\phi} {}^{,b} \nonumber \\ 
&& -  \tfrac{1}{4} \kappa^2 h^{}{}_{ab} h^{c}{}_{c} \bar{\phi} {}^{,a} \bar{\phi} {}^{,b} + \tfrac{1}{8} \kappa^2 \nu h^{}{}_{ab} h^{c}{}_{c} \bar{\phi} {}^{,a} \bar{\phi} {}^{,b}\Bigr] \\
\label{s3}
    \tilde{S}_{3} &=& \intx{x} \Bigl[\frac{\kappa \lambda \delta \phi h^{a}{}_{a} \bar{\phi}^3}{12 k_0^2} -  \frac{\kappa \lambda \nu \delta \phi h^{a}{}_{a} \bar{\phi}^3}{24 k_0^2} + \frac{k_1 \kappa \nu \delta \phi h^{b}{}_{b} \bar{\phi}^2 \bar{\phi}{}^{,a}{}_{,a}}{4 k_0^2} + \frac{k_1 \kappa h^{b}{}_{b} \bar{\phi}^2 \bar{\phi}{}_{,a} \delta \phi{}^{,a}}{2 k_0^2} \nonumber \\ 
    && + \frac{k_1 \kappa \delta \phi h^{b}{}_{b} \bar{\phi} \bar{\phi}{}_{,a} \bar{\phi}{}^{,a}}{2 k_0^2} + \frac{k_1 \kappa \nu \delta \phi h^{b}{}_{b} \bar{\phi} \bar{\phi}{}_{,a} \bar{\phi}{}^{,a}}{4 k_0^2} -  \frac{k_1 \kappa h^{}{}_{ab} \bar{\phi}^2 \delta \phi{}^{,a} \bar{\phi}{}^{,b}}{k_0^2} -  \frac{k_1 \kappa \delta \phi h^{}{}_{ab} \bar{\phi} \bar{\phi}{}^{,a} \bar{\phi}{}^{,b}}{k_0^2}\Bigr] 
\end{eqnarray}

\begin{eqnarray}
\label{s4}
\tilde{S}_{4} &=& \intx{x} \Biggl[- \frac{\kappa^2 \lambda h^{}{}_{ab} h^{ab} \bar{\phi}^4}{96 k_0^2} + \frac{\kappa^2 \lambda h^{a}{}_{a} h^{b}{}_{b} \bar{\phi}^4}{192 k_0^2} -  \frac{\kappa^2 \lambda \nu \bar{\phi}^4 (\delta \phi)^2}{96 k_0^2} -  \frac{k_1 \kappa^2 h^{}{}_{bc} h^{bc} \bar{\phi}^2 \bar{\phi}{}_{,a} \bar{\phi}{}^{,a}}{8 k_0^2} \nonumber \\ 
&& + \frac{k_1 \kappa^2 \nu h^{}{}_{bc} h^{bc} \bar{\phi}^2 \bar{\phi}{}_{,a} \bar{\phi}{}^{,a}}{16 k_0^2} + \frac{k_1 \kappa^2 h^{b}{}_{b} h^{c}{}_{c} \bar{\phi}^2 \bar{\phi}{}_{,a} \bar{\phi}{}^{,a}}{16 k_0^2} -  \frac{k_1 \kappa^2 \nu h^{b}{}_{b} h^{c}{}_{c} \bar{\phi}^2 \bar{\phi}{}_{,a} \bar{\phi}{}^{,a}}{32 k_0^2}  \nonumber \\ 
&& -  \frac{k_1 \kappa^2 \nu \bar{\phi}^2 (\delta \phi)^2 \bar{\phi}{}_{,a} \bar{\phi}{}^{,a}}{16 k_0^2} + \frac{k_1 \kappa^2 h^{}{}_{a}{}^{c} h^{}{}_{bc} \bar{\phi}^2 \bar{\phi}{}^{,a} \bar{\phi}{}^{,b}}{2 k_0^2} -  \frac{k_1 \kappa^2 \nu h^{}{}_{a}{}^{c} h^{}{}_{bc} \bar{\phi}^2 \bar{\phi}{}^{,a} \bar{\phi}{}^{,b}}{4 k_0^2} \nonumber \\
&& -  \frac{k_1 \kappa^2 h^{}{}_{ab} h^{c}{}_{c} \bar{\phi}^2 \bar{\phi}{}^{,a} \bar{\phi}{}^{,b}}{4 k_0^2} + \frac{k_1 \kappa^2 \nu h^{}{}_{ab} h^{c}{}_{c} \bar{\phi}^2 \bar{\phi}{}^{,a} \bar{\phi}{}^{,b}}{8 k_0^2}\Biggr]
\end{eqnarray}
Here, the indices $(a,b,c,\dots)$ and $(\mu,\nu,\rho,\dots)$ are used interchangeably to denote the gauge indices. 
$\tilde{S}_0$ leads to the well known free theory propagators for gravity and massive scalar field and the ghost field respectively,
\begin{eqnarray}
    \label{prop}
    D(x,x') &=&\intp{k}e^{ik\cdot(x-x')}D(k) = \langle\delta\phi(x)\delta\phi(x')\rangle; \nonumber \\
    D_{\alpha\beta\mu\nu}(x,x') & =& \intp{k} e^{ik\cdot(x-x')}D_{\alpha\beta\mu\nu}(k) = \langle h_{\alpha\beta}(x)h_{\mu\nu}(x')\rangle ; \\
    D^{G}_{\mu\nu}(x,x') &=& \intp{k}e^{ik\cdot(x-x')}D^{G}_{\mu\nu}(k) = \langle\bar{c}_{\mu}(x)c_{\nu}(x')\rangle \nonumber ;
\end{eqnarray}
where,
\begin{eqnarray}
    \label{props}
    D(k) &=& \frac{1}{k^2+\frac{m^2}{k_0}}; \\
    \label{propg}
    D_{\alpha\beta\mu\nu}(k) &=&\frac{\delta_{\alpha\mu}\delta_{\beta\nu}+\delta_{\alpha\nu}\delta_{\beta\mu} -\delta_{\alpha\beta}\delta_{\mu\nu}}{2 k^2} + (\alpha-1)\frac{\delta_{\alpha\mu}k_\beta k_\nu+\delta_{\alpha\nu}k_\beta k_\mu+\delta_{\beta\mu}k_\alpha k_\nu+\delta_{\beta\nu}k_\alpha k_\mu}{2k^4}; \\
    \label{propgh}
    D^{G}_{\mu\nu}(k) &=& \dfrac{1}{k^2}\delta_{\mu\nu}.
\end{eqnarray}
Looking at the structure of rest of the terms $\tilde{S}_i$, it is straightforward to conclude that all terms with odd combinations of $h_{\mu\nu}(x)$ and $\delta\phi(x)$ appearing in Eqs. (\ref{aeaf7}) will not contribute to the effective action, since $\langle h_{\alpha\beta}(x)\delta\phi(x')\rangle = 0$. Therefore, $\langle \tilde{S}_{1}\rangle = 0$ and there is no contribution at $\mathcal{O}(\bar{\phi})$ to the one-loop effective action. Similarly, $\langle \tilde{S}_{3}\rangle = \langle \tilde{S}_{1}\tilde{S}_{2}\rangle = \langle \tilde{S}_{1}^3\rangle = 0$, and hence at $\mathcal{O}(\bar{\phi}^{3})$ too, there is no contribution to the effective action. Hence, the only non-zero contributions in Eq. (\ref{aeaf7}) come at $\mathcal{O}(\bar{\phi}^{2})$ and $\mathcal{O}(\bar{\phi}^{4})$. In the latter, we ignore $\langle S_{1}^{4}\rangle$ terms since they are relevant at $\mathcal{O}(\kappa^4)$ and above while we are interested in terms up to $\kappa^2$ order. Expectation value of $\tilde{S}_{i}$ consists of local terms, and thus describes contributions from tadpole diagrams. 

The ghost term appears only in $\tilde{S}_{2}$. However, it can be shown that at $\mathcal{O}(\bar{\phi}^{2})$ it yields no nontrivial contributions, and as a result, has usually been ignored in past literature where only quadratic order corrections were considered \cite{mackay2010,saltas2017,bounakis2018}. Consider the ghost propagator (\ref{propgh}). Because there is no physical scale involved, the term containing ghost in (\ref{s2}) yields,
\begin{eqnarray}
    \label{loop00}
    \left\langle \intx{x}\omega c^{a} \bar{c}^{b} \bar{\phi} {}_{,a} \bar{\phi} {}_{,b} \right\rangle &=& \intx{x}\omega\bar{\phi} {}_{,a} \bar{\phi} {}_{,b} \langle c^{a} \bar{c}^{b} \rangle \nonumber \\
    &=& \intx{x}\omega\bar{\phi} {}_{,a} \bar{\phi} {}_{,b} \intp{k}\delta^{ab}\dfrac{1}{k^2},
\end{eqnarray}
which in four dimensions gives no physical result. The only nontrivial ghost contribution comes at quartic order in background field. 

Eventually, finding the one-loop corrections then boils down to evaluating up to $\kappa^{2}$ order, the quadratic and quartic order corrections from the following:
\begin{equation}
    \label{loop01}
    \Gamma = \langle\tilde{S}_{2}\rangle - \dfrac{1}{2}\langle\tilde{S}_{1}^{2}\rangle + \langle\tilde{S}_{4}\rangle - \langle\tilde{S}_{1}\tilde{S}_{3}\rangle + \dfrac{1}{2}\langle\tilde{S}_{1}^{2}\tilde{S}_{2}\rangle -\dfrac{1}{2}\langle\tilde{S}_{2}^{2}\rangle .
\end{equation}
In principle, solving Eq. (\ref{loop01}) broadly consists of two steps: (i) writing each term in terms of the Fourier space integral(s) of Green's functions found in Eqs. (\ref{props}) - (\ref{propgh}); and (ii) solving the resulting loop integrals. In this section, we restrict ourselves to writing just the divergent part of effective action, since there are already several thousand terms to deal with and writing their finite parts would introduce unnecessary complexity. We do consider finite part in the subsequent section, where we evaluate the effective potential after assuming all derivatives of background fields to be zero. 

\subsubsection{Calculating $\langle \tilde{S}_{i} \rangle$}
We first deal with $\langle\tilde{S}_{2}\rangle$ and $\langle\tilde{S}_{4}\rangle$. For convenience, we do not explicitly write the tensor indices of correlators, fields and their coefficients. First, the derivatives of field fluctuations are transformed to momentum space:
\begin{eqnarray}
\label{loop02}
\int d^{4}x A(x) \langle\partial^{m}\delta(x) \ \partial^{n}\delta(x')\rangle \longrightarrow \int d^{4}x \dfrac{d^{4}p}{(2\pi^{4})} A(x) (-ip)^{m} (ip)^{n} \langle \delta_{p}(x)\delta_{p}(x')\rangle,
\end{eqnarray}
where, $\delta (x)$ and $A(x)$ represent the field fluctuations and coefficients respectively. $\delta(x)$ here represents any of the fields ($\delta\phi(x),h_{\mu\nu}(x),c_{\mu}(x),\bar{c}_{\mu}(x)$), and is not to be confused with the Dirac delta function $\delta(x,x')$. $\langle \delta_{p}\delta_{p}\rangle$ represents the propagator(s) in momentum space. Then, $\langle \delta_{p}\delta_{p}\rangle$ is replaced with values of Green's function to obtain the loop integrals. For solving integrals here, we primarily use the results in Ref. \cite{bardin1999} to evaluate the divergent terms in dimensional regularization, except for some higher rank two-point integrals that appear below, which we solve by hand using well known prescriptions \cite{romao2019,george1975}. There are three types of loop integrals coming from Eqs. (\ref{s2}) and (\ref{s4}):
\begin{eqnarray}
\label{loop03}
\int d^{4}x \dfrac{d^{4}p}{(2\pi^{4})} A(x)\dfrac{1}{p^{2}} ; \quad \int d^{4}x \dfrac{d^{4}p}{(2\pi^{4})} A(x)\dfrac{p^{\mu}p^{\nu}}{p^{4}}; \quad  \int d^{4}x \dfrac{d^{4}p}{(2\pi^{4})} A(x)\dfrac{1}{p^{2}+\frac{m^2}{k_{0}}}.
\end{eqnarray}
The first two integrals are poleless, and vanish due to the lack of a physical scale \cite{bardin1999}. The third integral is straightforward and contributes to the divergent part. 
See appendix \ref{a1} for values of all integrals appearing here, including finite parts for some integrals used in the next section.

\subsubsection{Calculating $\langle \tilde{S}_{i}\tilde{S}_{j} \rangle$}
$\langle \tilde{S}_{1}\tilde{S}_{1} \rangle$ and $\langle \tilde{S}_{1}\tilde{S}_{3} \rangle$ contain terms of the form,
\begin{eqnarray}
    \label{loop04}
    & \intx{x}\intx{x'} A(x)B(x')\langle\delta\phi(x)\delta\phi(x')\rangle \langle h(x)h(x')\rangle \\
    =&\intx{x}\intp{k}\frac{d^4 k'}{2\pi^4}\frac{d^4 k''}{2\pi^4} A(x)\tilde{B}(k'') D_{\phi\phi}(k)D_{hh}(k') e^{-i(k+k')\cdot x}\delta^{(4)}(k+k'-k'') \nonumber \\
    \label{loop05}
    =& \intx{x}A(x)\intp{k}e^{-ik\cdot x}\tilde{B}(k) \intp{k'} D_{\phi\phi}(k-k')D_{hh}(k')
\end{eqnarray}
where $A(x), B(x')$ are classical coefficients, and $D_{\phi\phi}, D_{hh}$ are scalar and gravity propagators respectively; $\tilde{B}(k)$ is the Fourier transform of $B(x')$. There are also the derivatives of Eq. (\ref{loop04}) present, and are dealt with in a way similar to Eq. (\ref{loop02}), leading to factors of $k'^{\mu}$ in the loop integrals. Consequently, we encounter three types of loop integrals:
\begin{eqnarray}
    \label{loop06}
    \intp{k'} \dfrac{k'^{a}\dots k'^{b}}{(k'-k)^2 + \frac{m^2}{k_0}}; \quad \intp{k'} \dfrac{k'^{a}\dots k'^{b}}{k'^{2} ((k'-k)^2 + \frac{m^2}{k_0})}; \quad \intp{k'} \dfrac{k'^{a}\dots k'^{b}}{k'^{4} ((k'-k)^2 + \frac{m^2}{k_0})};
\end{eqnarray}
which constitute standard one-, two- and three-point $n$-rank integrals ($n=0,1,2$). 

Likewise, $\langle \tilde{S}_{2}\tilde{S}_{2} \rangle$ yields $4-$point correlators given by,
\begin{eqnarray}
    \label{loop07}
    \langle\delta\phi(x)\delta\phi(x)\delta\phi(x')\delta\phi(x')\rangle; \ \langle\delta\phi(x)\delta\phi(x)h(x')h(x')\rangle; \ \langle\delta\phi(x)\delta\phi(x)\bar{c}(x') c(x')\rangle; \nonumber \\
    \langle h(x) h(x)\bar{c}(x') c(x')\rangle; \ \langle \bar{c}(x) c(x) \bar{c}(x') c(x')\rangle; \ \langle h(x) h(x) h(x') h(x')\rangle
\end{eqnarray}
The second, third and fourth terms in (\ref{loop07}) are of the form $\langle \delta(x)\delta(x)\delta'(x')\delta'(x')\rangle$ (again, $\delta(x),\delta(x')$ denote the fields), thereby corresponding to disconnected tadpoles and hence do not give any meaningful contribution. The rest of $4-$point correlators in Eq. (\ref{loop07}) are resolved into $2-$point functions using Wick theorem \cite{peskin1995,schwartz2013}. Fortunately, the last term involving only graviton propagators can be ignored since it only contains $\mathcal{O}(\kappa^4)$ terms. Moreover, $\langle \bar{c}(x)\bar{c}(x')\rangle = \langle c(x)c(x')\rangle = 0$. Therefore, after applying Wick theorem, the final contribution in Eq. (\ref{loop07}) comes from,
\begin{eqnarray}
    \label{loop08a}
    \langle\delta\phi(x)\delta\phi(x)\delta\phi(x')\delta\phi(x')\rangle &=& \langle\delta\phi(x)\delta\phi(x')\rangle\langle\delta\phi(x)\delta\phi(x')\rangle + \langle\delta\phi(x)\delta\phi(x')\rangle\langle\delta\phi(x)\delta\phi(x')\rangle; \\
    \label{loop08b}
    \langle \bar{c}(x) c(x) \bar{c}(x') c(x')\rangle &=& \langle \bar{c}(x) c(x')\rangle \langle \bar{c}(x) c(x')\rangle.
\end{eqnarray}
Using Eq. (\ref{loop08a}) and (\ref{loop08b}) in $\langle \tilde{S}_{2}\tilde{S}_{2} \rangle$ along with Eqs. (\ref{props})-(\ref{propgh}), and Fourier transforming according to Eq. (\ref{loop02}) gives rise to up to rank-4 two-point integrals:
\begin{eqnarray}
    \label{loop09}
    \intp{k'} \dfrac{k'^{a}\dots k'^{b}}{(k'^{2}+\frac{m^2}{k_0}) ((k'-k)^2 + \frac{m^2}{k_0})};
\end{eqnarray}

\subsubsection{Calculating $\langle \tilde{S}_{i}\tilde{S}_{j}\tilde{S}_{k} \rangle$}
The last term to be evaluated is $\langle \tilde{S}_{1}\tilde{S}_{1}\tilde{S}_{2} \rangle$. It consists of six-point correlators given by,
\begin{eqnarray}
    \label{loop10}
    &\langle h(x)h(x'')\delta\phi(x)\delta\phi(x'')\bar{c}(x') c(x') \rangle ; \langle h(x)h(x'')\delta\phi(x)\delta\phi(x'')\delta\phi(x')\delta\phi(x') \rangle ; & \nonumber \\ 
    &\langle h(x)h(x'')\delta\phi(x)\delta\phi(x'') h(x')h(x') \rangle &
\end{eqnarray}
Again, the last term can be ignored since it has no terms up to $\mathcal{O}(\kappa^2)$. And the first term can be written as $\langle h(x)h(x'')\delta\phi(x)\delta\phi(x'')\rangle\langle\bar{c}(x') c(x') \rangle$, which implies disconnected diagrams and thus can also be ignored. So, ghost terms only end up in $\langle \tilde{S}_{2}\tilde{S}_{2} \rangle$. Hence, only the second term needs to be evaluated, which after applying Wick theorem similar to Eq. (\ref{loop08a}) turns out to be,
\begin{eqnarray}
    \label{loop11}
    \langle h(x)h(x'')\delta\phi(x)\delta\phi(x'')\delta\phi(x')\delta\phi(x') \rangle = & \langle h(x)h(x'')\rangle\langle\delta\phi(x)\delta\phi(x')\rangle\langle\delta\phi(x')\delta\phi(x'') \rangle \nonumber \\
    & + \langle h(x)h(x'')\rangle\langle\delta\phi(x)\delta\phi(x')\rangle\langle\delta\phi(x')\delta\phi(x'') \rangle ,
\end{eqnarray}
A typical scalar integral in $\langle \tilde{S}_{1}\tilde{S}_{1}\tilde{S}_{2} \rangle$ takes the form,
\begin{eqnarray}
    \label{loop12}
    &&\intx{x}\intx{x'}\intx{x''}\intp{k}\intp{k'}\intp{k''} A(x)B(x')C(x'')\times \nonumber \\ &&
    e^{-ik\cdot(x'-x)}e^{-ik''\cdot(x-x'')}e^{-ik'\cdot (x''-x')}D_{\phi\phi}(k)D_{\phi\phi}(k')D_{hh}(k'') \nonumber\\
   & =& \intx{x}\intp{p}\intp{k} A(x)\tilde{B}(p)e^{-ip\cdot x} \tilde{C}(k)e^{-ik\cdot x}\intp{k'} \times \nonumber \\ && D_{\phi\phi}(k'-p-k) D_{\phi\phi}(k'-k)D_{hh}(k'),
\end{eqnarray}
resulting in scalar and tensor two-, three- and four-point integrals:
\begin{eqnarray}
    \label{s112}
    \intp{k'}\dfrac{k'^{a}\dots k'^{b}}{d_{0}d_{1}d_{2}d_{3}}; \intp{k'}\dfrac{k'^{a}\dots k'^{b}}{d_{0}d_{1}d_{2}}; \intp{k'}\dfrac{k'^{a}\dots k'^{b}}{d_{0}d_{1}}.
\end{eqnarray}
where,
\begin{eqnarray}
    \label{loop13}
    d_{0} = (k'-k)^2 + \frac{m^2}{k_0}; \ d_{1} = (k'-k-p)^{2} + \frac{m^2}{k_0}; \ d_{2} = d_{3} = k'^2.
\end{eqnarray}
There are up to rank-3 four-point integrals in $\langle \tilde{S}_{1}\tilde{S}_{1}\tilde{S}_{2} \rangle$, and hence have no divergent part \cite{bardin1999}.

\subsubsection{Divergent part}
In total, there are several thousand terms that eventually add up to give the divergent part of Eq. (\ref{loop01}). After solving all the above integrals and extracting their divergent parts using dimensional regularization, we end up with Fourier transforms $\tilde{B}(k)$ (and $\tilde{C}(p)$ in case of six-point functions) with or without factors of $k^{a}$ and/or $p^{a}$. These expressions are transformed back to coordinate space as follows:
\begin{eqnarray}
    \label{divp0}
    \intx{x}\intp{p}\intp{k}A(x)\tilde{B}(p)\tilde{C}(k)e^{-ip\cdot x}e^{-ik\cdot x} k^{a}\dots k^{b} p^{\mu}\dots p^{\nu} \to \nonumber \\ \intx{x}(i\partial^{\mu})\dots(i\partial^{\nu})B(x) (i\partial^{a})\dots(i\partial^{b})C(x)
\end{eqnarray}
and likewise for other cases including $\langle \tilde{S}_{i}\tilde{S}_{j} \rangle$ and $\langle \tilde{S}_{i}\rangle$. Substituting these results for the divergent part in Eq. (\ref{loop01}), we get,
\begin{eqnarray}
    \label{divp1}
    divp(\Gamma) &=& \intx{x} L \Bigg[\frac{ k_{1} m^4 \bar{\phi}^2}{2 k_{0}^4} + \frac{3 m^4 \kappa^2 \bar{\phi}^2}{4 k_{0}^2} -  \frac{ m^2 \lambda \bar{\phi}^2}{4 k_{0}^3} -  \frac{5 m^4 \kappa^2 \nu \bar{\phi}^2}{8 k_{0}^2} \nonumber \\ 
    && + \frac{3 m^4 \kappa^2 \nu^2 \bar{\phi}^2}{16 k_{0}^2} + \frac{ k_{1} m^2 \bar{\phi} \bar{\phi}{}^{,a}{}_{,a}}{2 k_{0}^3} -  \frac{3 m^2 \kappa^2 \bar{\phi} \bar{\phi}{}^{,a}{}_{,a}}{4 k_{0}} + \frac{17 m^2 \kappa^2 \nu \bar{\phi} \bar{\phi}{}^{,a}{}_{,a}}{16 k_{0}} \nonumber \\ 
    && -  \frac{3 m^2 \kappa^2 \nu^2 \bar{\phi} \bar{\phi}{}^{,a}{}_{,a}}{8 k_{0}} + \frac{ m^2 \kappa^2 \omega \bar{\phi} \bar{\phi}{}^{,a}{}_{,a}}{4 k_{0}} + \frac{ m^2 \kappa^2 \nu \omega \bar{\phi} \bar{\phi}{}^{,a}{}_{,a}}{8 k_{0}} -  \tfrac{3}{8} \kappa^2 \nu \bar{\phi} \bar{\phi}{}^{,a}{}_{,a}{}^{,b}{}_{,b} \nonumber \\ 
    && + \tfrac{3}{16} \kappa^2 \nu^2 \bar{\phi} \bar{\phi}{}^{,a}{}_{,a}{}^{,b}{}_{,b} -  \tfrac{1}{4} \kappa^2 \omega \bar{\phi} \bar{\phi}{}^{,a}{}_{,a}{}^{,b}{}_{,b} -  \tfrac{1}{8} \kappa^2 \nu \omega \bar{\phi} \bar{\phi}{}^{,a}{}_{,a}{}^{,b}{}_{,b} \nonumber \\
    && - \frac{3 k_{1}^2 m^4 \bar{\phi}^4}{256 k_{0}^6 \pi^2} -  \frac{3 k_{1} m^4 \kappa^2 \bar{\phi}^4}{32 k_{0}^4 \pi^2} + \frac{ k_{1} m^2 \lambda \bar{\phi}^4}{16 k_{0}^5 \pi^2} + \frac{ m^2 \kappa^2 \lambda \bar{\phi}^4}{32 k_{0}^3 \pi^2} \nonumber \\ 
    && -  \frac{ \lambda^2 \bar{\phi}^4}{128 k_{0}^4 \pi^2} + \frac{ k_{1} m^4 \kappa^2 \nu \bar{\phi}^4}{16 k_{0}^4 \pi^2} -  \frac{17 m^2 \kappa^2 \lambda \nu \bar{\phi}^4}{768 k_{0}^3 \pi^2} -  \frac{3 k_{1} m^4 \kappa^2 \nu^2 \bar{\phi}^4}{128 k_{0}^4 \pi^2} \nonumber \\ 
    && + \frac{ m^2 \kappa^2 \lambda \nu^2 \bar{\phi}^4}{128 k_{0}^3 \pi^2} -  \frac{19 k_{1}^2 m^2 \bar{\phi}^3 \bar{\phi} {}^{,a}{}_{,a}}{128 k_{0}^5 \pi^2} + \frac{ k_{1} \lambda \bar{\phi}^3 \bar{\phi} {}^{,a}{}_{,a}}{64 k_{0}^4 \pi^2} -  \frac{ \kappa^2 \lambda \bar{\phi}^3 \bar{\phi} {}^{,a}{}_{,a}}{64 k_{0}^2 \pi^2} \nonumber \\ 
    && -  \frac{ k_{1} m^2 \kappa^2 \nu \bar{\phi}^3 \bar{\phi} {}^{,a}{}_{,a}}{128 k_{0}^3 \pi^2} + \frac{3 \kappa^2 \lambda \nu \bar{\phi}^3 \bar{\phi} {}^{,a}{}_{,a}}{128 k_{0}^2 \pi^2} -  \frac{ \kappa^2 \lambda \nu^2 \bar{\phi}^3 \bar{\phi} {}^{,a}{}_{,a}}{128 k_{0}^2 \pi^2} + \frac{ k_{1} m^2 \kappa^2 \omega \bar{\phi}^3 \bar{\phi} {}^{,a}{}_{,a}}{64 k_{0}^3 \pi^2} \nonumber \\ 
    && -  \frac{ \kappa^2 \lambda \omega \bar{\phi}^3 \bar{\phi} {}^{,a}{}_{,a}}{192 k_{0}^2 \pi^2} -  \frac{ k_{1} m^2 \kappa^2 \nu \omega \bar{\phi}^3 \bar{\phi} {}^{,a}{}_{,a}}{128 k_{0}^3 \pi^2} + \frac{ \kappa^2 \lambda \nu \omega \bar{\phi}^3 \bar{\phi} {}^{,a}{}_{,a}}{384 k_{0}^2 \pi^2} -  \frac{11 k_{1}^2 m^2 \bar{\phi}^2 \bar{\phi} {}_{,a} \bar{\phi} {}^{,a}}{128 k_{0}^5 \pi^2} \nonumber \\ 
    && -  \frac{3 k_{1} m^2 \kappa^2 \bar{\phi}^2 \bar{\phi} {}_{,a} \bar{\phi} {}^{,a}}{32 k_{0}^3 \pi^2} -  \frac{3 k_{1} \lambda \bar{\phi}^2 \bar{\phi} {}_{,a} \bar{\phi} {}^{,a}}{64 k_{0}^4 \pi^2} + \frac{5 k_{1} m^2 \kappa^2 \nu \bar{\phi}^2 \bar{\phi} {}_{,a} \bar{\phi} {}^{,a}}{32 k_{0}^3 \pi^2} + \frac{ \kappa^2 \lambda \nu \bar{\phi}^2 \bar{\phi} {}_{,a} \bar{\phi} {}^{,a}}{256 k_{0}^2 \pi^2} \nonumber \\ 
    && -  \frac{3 k_{1} m^2 \kappa^2 \nu^2 \bar{\phi}^2 \bar{\phi} {}_{,a} \bar{\phi} {}^{,a}}{64 k_{0}^3 \pi^2} + \frac{3 k_{1} m^2 \kappa^2 \omega \bar{\phi}^2 \bar{\phi} {}_{,a} \bar{\phi} {}^{,a}}{64 k_{0}^3 \pi^2} -  \frac{ \kappa^2 \lambda \omega \bar{\phi}^2 \bar{\phi} {}_{,a} \bar{\phi} {}^{,a}}{32 k_{0}^2 \pi^2} + \frac{ k_{1} m^2 \kappa^2 \nu \omega \bar{\phi}^2 \bar{\phi} {}_{,a} \bar{\phi} {}^{,a}}{128 k_{0}^3 \pi^2} \nonumber \\ 
    && + \frac{7 k_{1}^2 \bar{\phi}^2 \bar{\phi} {}^{,b}{}_{,b}{}_{,a} \bar{\phi} {}^{,a}}{128 k_{0}^4 \pi^2} + \frac{5 k_{1} \kappa^2 \omega \bar{\phi}^2 \bar{\phi} {}^{,b}{}_{,b}{}_{,a} \bar{\phi} {}^{,a}}{384 k_{0}^2 \pi^2} -  \frac{ k_{1} \kappa^2 \nu \omega \bar{\phi}^2 \bar{\phi} {}^{,b}{}_{,b}{}_{,a} \bar{\phi} {}^{,a}}{256 k_{0}^2 \pi^2} -  \frac{ k_{1}^2 \bar{\phi}^2 \bar{\phi} {}^{,a} \bar{\phi} {}^{,b}{}_{,a}{}_{,b}}{24 k_{0}^4 \pi^2} \nonumber \\ 
    && -  \frac{3 k_{1} \kappa^2 \omega \bar{\phi}^2 \bar{\phi} {}^{,a} \bar{\phi} {}^{,b}{}_{,a}{}_{,b}}{128 k_{0}^2 \pi^2} -  \frac{ k_{1}^2 \bar{\phi}^2 \bar{\phi} {}^{,a}{}_{,a} \bar{\phi} {}^{,b}{}_{,b}}{256 k_{0}^4 \pi^2} + \frac{ k_{1} \kappa^2 \bar{\phi}^2 \bar{\phi} {}^{,a}{}_{,a} \bar{\phi} {}^{,b}{}_{,b}}{32 k_{0}^2 \pi^2} -  \frac{3 k_{1} \kappa^2 \nu \bar{\phi}^2 \bar{\phi} {}^{,a}{}_{,a} \bar{\phi} {}^{,b}{}_{,b}}{64 k_{0}^2 \pi^2} \nonumber \\ 
    && + \frac{3 k_{1} \kappa^2 \nu^2 \bar{\phi}^2 \bar{\phi} {}^{,a}{}_{,a} \bar{\phi} {}^{,b}{}_{,b}}{128 k_{0}^2 \pi^2} + \frac{ k_{1} \kappa^2 \omega \bar{\phi}^2 \bar{\phi} {}^{,a}{}_{,a} \bar{\phi} {}^{,b}{}_{,b}}{64 k_{0}^2 \pi^2} -  \frac{ k_{1} \kappa^2 \nu \omega \bar{\phi}^2 \bar{\phi} {}^{,a}{}_{,a} \bar{\phi} {}^{,b}{}_{,b}}{128 k_{0}^2 \pi^2} + \frac{ k_{1}^2 \bar{\phi} \bar{\phi} {}_{,a} \bar{\phi} {}^{,a} \bar{\phi} {}^{,b}{}_{,b}}{16 k_{0}^4 \pi^2} \nonumber \\ 
    && + \frac{ k_{1} \kappa^2 \bar{\phi} \bar{\phi} {}_{,a} \bar{\phi} {}^{,a} \bar{\phi} {}^{,b}{}_{,b}}{16 k_{0}^2 \pi^2} -  \frac{25 k_{1} \kappa^2 \nu \bar{\phi} \bar{\phi} {}_{,a} \bar{\phi} {}^{,a} \bar{\phi} {}^{,b}{}_{,b}}{256 k_{0}^2 \pi^2} + \frac{3 k_{1} \kappa^2 \nu^2 \bar{\phi} \bar{\phi} {}_{,a} \bar{\phi} {}^{,a} \bar{\phi} {}^{,b}{}_{,b}}{64 k_{0}^2 \pi^2} + \frac{3 k_{1} \kappa^2 \omega \bar{\phi} \bar{\phi} {}_{,a} \bar{\phi} {}^{,a} \bar{\phi} {}^{,b}{}_{,b}}{32 k_{0}^2 \pi^2} \nonumber \\ 
    && -  \frac{ k_{1} \kappa^2 \nu \omega \bar{\phi} \bar{\phi} {}_{,a} \bar{\phi} {}^{,a} \bar{\phi} {}^{,b}{}_{,b}}{64 k_{0}^2 \pi^2} + \frac{ k_{1}^2 \bar{\phi}^2 \bar{\phi} {}^{,a} \bar{\phi} {}_{,a}{}^{,b}{}_{,b}}{384 k_{0}^4 \pi^2} -  \frac{ k_{1} \kappa^2 \omega \bar{\phi}^2 \bar{\phi} {}^{,a} \bar{\phi} {}_{,a}{}^{,b}{}_{,b}}{192 k_{0}^2 \pi^2} -  \frac{ k_{1} \kappa^2 \nu \omega \bar{\phi}^2 \bar{\phi} {}^{,a} \bar{\phi} {}_{,a}{}^{,b}{}_{,b}}{256 k_{0}^2 \pi^2} \nonumber \\ 
    && -  \frac{ k_{1}^2 \bar{\phi}^3 \bar{\phi} {}^{,a}{}_{,a}{}^{,b}{}_{,b}}{256 k_{0}^4 \pi^2} + \frac{ k_{1} \kappa^2 \nu \bar{\phi} {}_{,a} \bar{\phi} {}^{,a} \bar{\phi} {}_{,b} \bar{\phi} {}^{,b}}{256 k_{0}^2 \pi^2} -  \frac{ k_{1} \kappa^2 \bar{\phi} \bar{\phi} {}^{,a} \bar{\phi} {}_{,a}{}_{,b} \bar{\phi} {}^{,b}}{16 k_{0}^2 \pi^2} + \frac{ k_{1} \kappa^2 \nu \bar{\phi} \bar{\phi} {}^{,a} \bar{\phi} {}_{,a}{}_{,b} \bar{\phi} {}^{,b}}{128 k_{0}^2 \pi^2} \nonumber \\ 
    && -  \frac{ k_{1} \kappa^2 \omega \bar{\phi} \bar{\phi} {}^{,a} \bar{\phi} {}_{,a}{}_{,b} \bar{\phi} {}^{,b}}{16 k_{0}^2 \pi^2} -  \frac{ k_{1}^2 \bar{\phi}^2 \bar{\phi} {}_{,a}{}_{,b} \bar{\phi} {}^{,a}{}^{,b}}{128 k_{0}^4 \pi^2} -  \frac{ k_{1} \kappa^2 \bar{\phi}^2 \bar{\phi} {}_{,a}{}_{,b} \bar{\phi} {}^{,a}{}^{,b}}{32 k_{0}^2 \pi^2} -  \frac{ k_{1} \kappa^2 \omega \bar{\phi}^2 \bar{\phi} {}_{,a}{}_{,b} \bar{\phi} {}^{,a}{}^{,b}}{32 k_{0}^2 \pi^2} \Bigg]
\end{eqnarray}
where, $L=-1/8\pi^{2}\epsilon$ ($\epsilon=n-4$) as the dimensionality $n\to 4$. 
As expected, there are no $\alpha$ dependent terms. Although not explicitly shown here, factors of $1/\alpha$ appear in individual pieces in Eq. (\ref{loop01}). However, when all contributions are added to evaluate $\Gamma$, these terms cancel so that the final result is gauge-invariant. Final result for divergent part of $\Gamma$ after removing bookkeeping parameters ($\omega\to 1$, $\nu \to 1$) in the Landau gauge ($\alpha \to 0$) leads to the covariant corrections:
\begin{eqnarray}
    \label{finalEA}
    divp(\Gamma) &=& \int d^4 x\Big[\frac{  k_{1} m^4  \bar{\phi}^2}{2  k_{0}^4} + \frac{5 m^4 \kappa^2  \bar{\phi}^2}{16  k_{0}^2} -  \frac{ m^2 \lambda  \bar{\phi}^2}{4  k_{0}^3} + \frac{  k_{1} m^2  \bar{\phi}  \bar{\phi} {}^{,a}{}_{,a}}{2  k_{0}^3} + \frac{5 m^2 \kappa^2  \bar{\phi}  \bar{\phi} {}^{,a}{}_{,a}}{16  k_{0}} \nonumber \\ 
    && -  \tfrac{9}{16} \kappa^2   \phi   \phi {}^{,a}{}_{,a}{}^{,b}{}_{,b}  - \frac{3  k_{1}^2 m^4 \bar{  \phi}^4}{32  k_{0}^6} -  \frac{7  k_{1} m^4 \kappa^2 \bar{  \phi}^4}{16  k_{0}^4} + \frac{  k_{1} m^2 \lambda \bar{  \phi}^4}{2  k_{0}^5} + \frac{13 m^2 \kappa^2 \lambda \bar{  \phi}^4}{96  k_{0}^3} \nonumber \\ 
    && -  \frac{ \lambda^2 \bar{  \phi}^4}{16  k_{0}^4} -  \frac{19  k_{1}^2 m^2 \bar{  \phi}^3 \bar{  \phi}{}^{,a}{}_{,a}}{16  k_{0}^5} + \frac{  k_{1} \lambda \bar{  \phi}^3 \bar{  \phi}{}^{,a}{}_{,a}}{8  k_{0}^4} -  \frac{ \kappa^2 \lambda \bar{  \phi}^3 \bar{  \phi}{}^{,a}{}_{,a}}{48  k_{0}^2} \nonumber \\ 
    && -  \frac{11  k_{1}^2 m^2 \bar{  \phi}^2 \bar{  \phi}{}_{,a} \bar{  \phi}{}^{,a}}{16  k_{0}^5} + \frac{9  k_{1} m^2 \kappa^2 \bar{  \phi}^2 \bar{  \phi}{}_{,a} \bar{  \phi}{}^{,a}}{16  k_{0}^3} -  \frac{3  k_{1} \lambda \bar{  \phi}^2 \bar{  \phi}{}_{,a} \bar{  \phi}{}^{,a}}{8  k_{0}^4} -  \frac{7 \kappa^2 \lambda \bar{  \phi}^2 \bar{  \phi}{}_{,a} \bar{  \phi}{}^{,a}}{32  k_{0}^2} \nonumber \\ 
    && + \frac{7  k_{1}^2 \bar{  \phi}^2 \bar{  \phi}{}^{,b}{}_{,b}{}_{,a} \bar{  \phi}{}^{,a}}{16  k_{0}^4} + \frac{7  k_{1} \kappa^2 \bar{  \phi}^2 \bar{  \phi}{}^{,b}{}_{,b}{}_{,a} \bar{  \phi}{}^{,a}}{96  k_{0}^2} -  \frac{  k_{1}^2 \bar{  \phi}^2 \bar{  \phi}{}^{,a} \bar{  \phi}{}^{,b}{}_{,a}{}_{,b}}{3  k_{0}^4} -  \frac{3  k_{1} \kappa^2 \bar{  \phi}^2 \bar{  \phi}{}^{,a} \bar{  \phi}{}^{,b}{}_{,a}{}_{,b}}{16  k_{0}^2} \nonumber \\ 
    && -  \frac{  k_{1}^2 \bar{  \phi}^2 \bar{  \phi}{}^{,a}{}_{,a} \bar{  \phi}{}^{,b}{}_{,b}}{32  k_{0}^4} + \frac{  k_{1} \kappa^2 \bar{  \phi}^2 \bar{  \phi}{}^{,a}{}_{,a} \bar{  \phi}{}^{,b}{}_{,b}}{8  k_{0}^2} + \frac{  k_{1}^2 \bar{  \phi} \bar{  \phi}{}_{,a} \bar{  \phi}{}^{,a} \bar{  \phi}{}^{,b}{}_{,b}}{2  k_{0}^4} + \frac{23  k_{1} \kappa^2 \bar{  \phi} \bar{  \phi}{}_{,a} \bar{  \phi}{}^{,a} \bar{  \phi}{}^{,b}{}_{,b}}{32  k_{0}^2} \nonumber \\ 
    && + \frac{  k_{1}^2 \bar{  \phi}^2 \bar{  \phi}{}^{,a} \bar{  \phi}{}_{,a}{}^{,b}{}_{,b}}{48  k_{0}^4} -  \frac{7  k_{1} \kappa^2 \bar{  \phi}^2 \bar{  \phi}{}^{,a} \bar{  \phi}{}_{,a}{}^{,b}{}_{,b}}{96  k_{0}^2} -  \frac{  k_{1}^2 \bar{  \phi}^3 \bar{  \phi}{}^{,a}{}_{,a}{}^{,b}{}_{,b}}{32  k_{0}^4} + \frac{  k_{1} \kappa^2 \bar{  \phi}{}_{,a} \bar{  \phi}{}^{,a} \bar{  \phi}{}_{,b} \bar{  \phi}{}^{,b}}{32  k_{0}^2} \nonumber \\ 
    && -  \frac{15  k_{1} \kappa^2 \bar{  \phi} \bar{  \phi}{}^{,a} \bar{  \phi}{}_{,a}{}_{,b} \bar{  \phi}{}^{,b}}{16  k_{0}^2} -  \frac{  k_{1}^2 \bar{  \phi}^2 \bar{  \phi}{}_{,a}{}_{,b} \bar{  \phi}{}^{,a}{}^{,b}}{16  k_{0}^4} -  \frac{  k_{1} \kappa^2 \bar{  \phi}^2 \bar{  \phi}{}_{,a}{}_{,b} \bar{  \phi}{}^{,a}{}^{,b}}{2  k_{0}^2}\Big]
\end{eqnarray} 
If instead we turn off the DV connections by setting $\nu = 0$ and choose $\alpha = 1, \omega = 1$, we recover gauge-dependent results obtained in the past by Steinwachs and Kamenshchik \cite{steinwachs2011}, where they calculated the one-loop divergences for a general scalar-tensor theory that in the single field limit (with the identifications $U=1, G=K$, and $V=V/\gamma^4$ in their notations) encompasses the model (\ref{action}). Similarly, in the case $k_1 = 0, k_0 = 1$ we recover the gauge-invariant calculations of Mackay and Toms \cite{mackay2010} (excluding cosmological constant and nonminimal coupling to gravity).

\subsection{Renormalization and Comparisons}
Not all the divergences in Eq. (\ref{finalEA}) can be absorbed by renormalizing the parameters in the classical action (\ref{action}), particularly the quartic derivatives of $\bar{\phi}(x)$, which are absent in the classical action. However, we need not worry about these UV divergences since the current framework is an effective theory approach, and we assume that such divergences are resolved by some high energy theory. For now, we consider only the counterterms for terms present in the classical action functional so as to absorb corresponding divergent parts, which will in turn induce 1-loop corrections to the parameters $\frac{m^2}{k_0}, \frac{k_1}{k_0^2}, \frac{\lambda}{k_0^2}$ of the theory (\ref{action}).

We start by re-writing Eq. (\ref{finalEA}) in the form,
\begin{eqnarray}
    \label{ren00}
    divp(\Gamma) = L \intx{x} (A \bar{\phi}\Box\bar{\phi} + B\bar{\phi}^2 + C \bar{\phi}^4 + D \bar{\phi}^2 \partial_{\mu}\bar{\phi}\partial^{\mu}\bar{\phi})
\end{eqnarray}
where we have ignored the terms not present in the classical background action. We note that the terms of the form $\bar{\phi}^{3}\Box\bar{\phi}$ in Eq. (\ref{finalEA}) are transformed to $-3\bar{\phi}^2\partial_{\mu}\bar{\phi}\partial^{\mu}\bar{\phi}$ after by-parts integration. The coefficients $A,B,C,D$ are read off from Eq. (\ref{finalEA}):
\begin{eqnarray}
    \label{ren01}
    A &=& \dfrac{5m^{2}\kappa^{2}}{16 k_{0}} + \dfrac{k_{1} m^{2}}{2 k_{0}^{3}}; \nonumber \\
    B &=& \dfrac{k_{1}m^{4}}{2 k_{0}^{4}} + \dfrac{5 m^{4}\kappa^{2}}{16 k_{0}^{2}} - \dfrac{m^{2}\lambda}{4 k_{0}^{3}}; \nonumber \\
    C &=& - \dfrac{3 k_{1}^{2}m^{4}}{32 k_{0}^{6}} - \dfrac{7 k_{1}m^{4}\kappa^{2}}{16 k_{0}^{4}} + \dfrac{k_{1}m^{2}\lambda^2}{k_{0}^{5}} + \dfrac{13 m^2 \kappa^2 \lambda}{96 k_0^3} - \dfrac{\lambda^2}{16 k_0^4}; \nonumber \\
    D &=& \dfrac{23 k_1^2 m^2}{8 k_0^5} + \dfrac{9 k_1 m^2 \kappa^2}{16 k_0^3} - \dfrac{3 k_1 \lambda}{4 k_0^4} - \dfrac{5\kappa^2 \lambda}{32 k_0^2}.
\end{eqnarray}
Taking into account the field Renormalization $\bar{\phi}\to Z^{1/2}\bar{\phi}$, the classical background Lagrangian reads,
\begin{eqnarray}
    \label{ren02}
    \mathcal{L}_{Z} = -\dfrac{1}{2} Z \bar{\phi}\Box\bar{\phi} + \frac{1}{2}\dfrac{m^2}{k_0} Z \bar{\phi}^2 + \frac{\lambda}{24 k_0^2} Z^2 \bar{\phi}^4 + \frac{1}{2}\frac{k_1}{k_0^2} Z^2 \bar{\phi}^2 \partial_{\mu}\bar{\phi}\partial^{\mu}\bar{\phi}
\end{eqnarray}
Suppose, the renormalized Lagrangian is given in terms of renormalized parameters as follows,
\begin{eqnarray}
    \label{ren03}
    \mathcal{L}_{r} = -\dfrac{1}{2} \bar{\phi}\Box\bar{\phi} + \frac{1}{2}\left(\dfrac{m^2}{k_0}\right)_{r} \bar{\phi}^2 + \frac{1}{24}\left(\frac{\lambda}{k_0^2}\right)_{r} \bar{\phi}^4 + \frac{1}{2}\left(\frac{k_1}{k_0^2}\right)_{r} \bar{\phi}^2 \partial_{\mu}\bar{\phi}\partial^{\mu}\bar{\phi}
\end{eqnarray}
where $(\cdot)_{r}$ represents the renormalized parameter. The counterterm Lagrangian is then defined as $\delta\mathcal{L} = \mathcal{L}_{r} - \mathcal{L}_{Z}$. Accordingly, the counterterms for field and other parameters are as follows:
\begin{eqnarray}
    \label{ren04}
    \delta_{Z} = Z-1; & \quad \delta\left(\dfrac{m^2}{k_0}\right) = \dfrac{m^2}{k_0}Z - \left(\dfrac{m^2}{k_0}\right)_{r} ; \nonumber \\
    \delta\left(\dfrac{\lambda}{k_0^2}\right) = \dfrac{\lambda}{k_0^2} Z^2 - \left(\dfrac{\lambda}{k_0^2}\right)_{r}; & \quad \delta\left(\dfrac{k_1}{k_0^2}\right) = \dfrac{k_1}{k_0^2} Z^2 - \left(\dfrac{k_1}{k_0^2}\right)_{r}.
\end{eqnarray}
These counterterms are fixed by demanding that $divp(\Gamma) = -\intx{x}\delta\mathcal{L}$. With some algebraic manipulations, the counterterms read,
\begin{eqnarray}
    \label{ren05}
    \delta_{Z} = -\dfrac{A}{4\pi^2 \epsilon} ; & \quad \delta\left(\dfrac{m^2}{k_0}\right) = \dfrac{B}{4\pi^2 \epsilon} ; \nonumber \\
    \delta\left(\dfrac{\lambda}{k_0^2}\right) = \dfrac{3 C}{\pi^2 \epsilon}; & \quad \delta\left(\dfrac{k_1}{k_0^2}\right) = \dfrac{D}{4\pi^2\epsilon}. 
\end{eqnarray}
Using Eq. (\ref{ren05}) in (\ref{ren04}), we find the one-loop corrections to coupling parameters in terms of the coefficients $A,B,C,D$, 
\begin{eqnarray}
    \label{ren06}
    \Delta \left(\dfrac{m^2}{k_0}\right) = \dfrac{m^2 A}{4\pi^2 k_0 \epsilon} + \dfrac{B}{4\pi^2 \epsilon}; \nonumber \\
    \Delta\left(\dfrac{\lambda}{k_0^2}\right) = \dfrac{3 C}{\pi^2 \epsilon} + \dfrac{\lambda A}{2\pi^2 k_0^2 \epsilon}; \\
    \Delta \left(\dfrac{k_1}{k_0^2}\right) = \dfrac{k_1 A}{2\pi^2 k_0^2\epsilon} + \dfrac{D}{4\pi^2\epsilon}. \nonumber 
\end{eqnarray}
For the sake of comparisons, and also as a crosscheck, we point out that upon choosing $\nu=0,\alpha=1,\omega=0$ in the case $k_1 = 0, k_0 = 1$, the gauge-dependent one-loop quantum gravitational correction to $\phi^4$ theory first calculated by Rodigast and Schuster \cite{rodigast2010a} is recovered: $\Delta\lambda = \frac{\kappa^2}{4\pi^2\epsilon}(m^2\lambda - 3\lambda^2/4\kappa^2)$. Note that, all gravitational corrections in Eq. (\ref{ren06}) appear with a factor of $\kappa$, while the ones without it are nongravitational corrections that could in principle be obtained from flat space quantum field theory. Also, in the gauge covariant version of the same case (viz. $\nu=1,\alpha=0,\omega=1$ with $k_1 = 0, k_0 = 1$), our results match that of Pietrykowski \cite{artur2013}. 

In a similar spirit, we would like to shed some light on the extensions of the work of Ref. \cite{mackay2010}. There, a self-interacting scalar field with nonminimal coupling to gravity (of the form $\xi R \phi^2/2$) was considered and the corresponding field and mass renormalizations were studied. The action in Ref. \cite{mackay2010} matches ours if we put $k_1 = 0, k_0 = 1$ and add $\xi R \phi^2/2$. However, corrections to quartic coupling including contributions from the nonminimal coupling have not been calculated so far. Without going into the details, partly because the process is more or less unchanged, we present here the covariant one-loop corrections to quartic coupling $\lambda$ so as to complete the analysis of Ref. \cite{mackay2010}, 
\begin{eqnarray}
    \label{ren07}
    \delta \lambda = \dfrac{3\lambda^2}{16\pi^2\epsilon} + \dfrac{\kappa^2}{\pi^2\epsilon}\left(\dfrac{9}{16}m^2\lambda + \dfrac{21}{8}m^2\lambda\xi^2 - \dfrac{3}{2}m^2\lambda\xi\right). 
\end{eqnarray}

\section{\label{sec5}Effective potential}
It is evident from the analysis so far that extracting any more information, in the form of finite corrections for example, is a cumbersome task. A resolution to this problem lies in making a reasonable compromise, wherein the derivatives of background fields are ignored basis the assumption that either the background field is constant due to a symmetry or it is slowly varying. The resulting effective action is known as effective potential. One of the first instances of this workaround is the well known Coleman Weinberg potential \cite{coleman1973,weinberg1973}. This approximation holds up especially during inflation, where the slow-rolling condition requires fields to be slowly varying. In this section, we evaluate the effective potential of the theory (\ref{action}) including finite terms and infer cosmological implications. 

We begin by substituting $\partial^{\mu}\bar{\phi}=0$ in Eqs. (\ref{s1})-(\ref{s4}), resulting in,
\begin{eqnarray}
    \label{es1}
    \tilde{S}_{1} &=& \intx{x} \Bigl[\frac{m^2 \kappa \delta \phi h^{a}{}_{a} \bar{\phi}}{2 k_0} -  \frac{m^2 \kappa \nu \delta \phi h^{a}{}_{a} \bar{\phi}}{4 k_0} \Bigr]; \\
\label{es2}
\tilde{S}_{2} &=& \intx{x} \Bigl[ -\frac{m^2 \kappa^2 h^{}{}_{ab} h^{ab} \bar{\phi}^2}{8 k_0} + \frac{m^2 \kappa^2 h^{a}{}_{a} h^{b}{}_{b} \bar{\phi}^2}{16 k_0} + \frac{\lambda \bar{\phi}^2 (\delta \phi)^2}{4 k_0^2} -  \frac{m^2 \kappa^2 \nu \bar{\phi}^2 (\delta \phi)^2}{8 k_0} \nonumber \\ 
&& + \frac{k_1 \bar{\phi}^2 \delta \phi{}_{,a} \delta \phi{}^{,a}}{2 k_0^2} \Bigr]; \\
\label{es3}
    \tilde{S}_{3} &=& \intx{x} \Bigl[\frac{\kappa \lambda \delta \phi h^{a}{}_{a} \bar{\phi}^3}{12 k_0^2} -  \frac{\kappa \lambda \nu \delta \phi h^{a}{}_{a} \bar{\phi}^3}{24 k_0^2} \Bigr]; \\
\label{es4}
\tilde{S}_{4} &=& \intx{x} \Bigl[- \frac{\kappa^2 \lambda h^{}{}_{ab} h^{ab} \bar{\phi}^4}{96 k_0^2} + \frac{\kappa^2 \lambda h^{a}{}_{a} h^{b}{}_{b} \bar{\phi}^4}{192 k_0^2} -  \frac{\kappa^2 \lambda \nu \bar{\phi}^4 (\delta \phi)^2}{96 k_0^2} \Bigr].
\end{eqnarray}
Using the above expressions in Eq. (\ref{loop01}) and following the steps outlined in the Sec. \ref{loopint}, we obtain the covariant effective potential,
\begin{eqnarray}
    \label{efpot0}
\Gamma_{eff}[\bar{\phi}] &=& \dfrac{1}{8\pi^2}\intx{x} [A_{1}\frac{1}{\epsilon}\bar{\phi}^2 + A_{2}\bar{\phi}^2 + B_{1}\frac{1}{\epsilon}\bar{\phi}^4 + B_{2}\bar{\phi}^4]
\end{eqnarray} 
where, $A_1$ and $B_1$ are the same as $B$ and $C$ from Eq. (\ref{ren01}) respectively, and,  
\begin{eqnarray}
    \label{efpot1}
    A_{2} &=& (\gamma + \log(\pi)) (- \frac{ k_1 m^4}{4  k_0^4} -  \frac{5 m^4 \kappa^2}{32  k_0^2} + \frac{ m^2 \lambda}{8  k_0^3})  + \frac{3 k_1 m^4}{8  k_0^4} + \frac{ m^4 \kappa^2}{4  k_0^2} -  \frac{ m^2 \lambda}{8  k_0^3} \nonumber \\ && + (- \frac{ k_1 m^4}{4  k_0^4} -  \frac{5 m^4 \kappa^2}{32  k_0^2} + \frac{ m^2 \lambda}{8  k_0^3}) \log(\frac{m^2}{ k_0 \mu^2}) \nonumber \\ && - \frac{1}{\bar{\phi}} \intp{k} e^{-ik\cdot x} \tilde{\bar{\phi}} \left( \frac{3 m^4 \kappa^2 \log\bigl(1 + \frac{ k_0  k^2}{m^2}\bigr)}{32  k_0^2} -  \frac{3 m^6 \kappa^2 \log\bigl(1 + \frac{ k_0  k^2}{m^2}\bigr)}{32  k_0^3  k^2}\right); \\
    \label{b2}
    B_{2} &=& - \frac{9  k_1^2 m^4}{128  k_0^6} -  \frac{5  k_1 m^4 \kappa^2}{16  k_0^4} + \frac{  k_1 m^2 \lambda}{4  k_0^5} + \frac{25 m^2 \kappa^2 \lambda}{192  k_0^3} -  \frac{ \lambda^2}{16  k_0^4} \nonumber \\ && + (\gamma + \log(\pi)) (\frac{3  k_1^2 m^4}{64  k_0^6} + \frac{7  k_1 m^4 \kappa^2}{32  k_0^4} -  \frac{  k_1 m^2 \lambda}{4  k_0^5} -  \frac{13 m^2 \kappa^2 \lambda}{192  k_0^3} + \frac{ \lambda^2}{32  k_0^4}) \nonumber \\ && + (\frac{3  k_1^2 m^4}{64  k_0^6} + \frac{5  k_1 m^4 \kappa^2}{32  k_0^4} -  \frac{  k_1 m^2 \lambda}{8  k_0^5} -  \frac{13 m^2 \kappa^2 \lambda}{192  k_0^3} + \frac{ \lambda^2}{32  k_0^4}) \log(\frac{m^2}{ k_0 \mu^2}) \nonumber \\ && 
    - \frac{1}{\bar{\phi}^3}\intp{k} e^{-ik\cdot x} \tilde{\bar{\phi}^3} \Bigg( \frac{ m^2 \kappa^2 \lambda \log\bigl(1 + \frac{ k_0 k^2}{m^2}\bigr)}{32  k_0^3}\Bigg) \nonumber \\ && + \frac{1}{\bar{\phi}^2}\intp{k} e^{-ik\cdot x} \tilde{\bar{\phi}^2} \Bigg( -\frac{ m^2 \kappa^2 \lambda \log\Bigl(\frac{1 + \bigl(1 + \frac{4 m^2}{ k_0 k^2}\bigr)^{1/2}}{-1 + \bigl(1 + \frac{4 m^2}{ k_0 k^2}\bigr)^{1/2}}\Bigr) \bigl(1 + \frac{4 m^2}{ k_0 k^2}\bigr)^{1/2}}{32  k_0^3} + \nonumber \\ &&  \frac{ \lambda^2 \log\Bigl(\frac{1 + \bigl(1 + \frac{4 m^2}{ k_0 k^2}\bigr)^{1/2}}{-1 + \bigl(1 + \frac{4 m^2}{ k_0 k^2}\bigr)^{1/2}}\Bigr) \bigl(1 + \frac{4 m^2}{ k_0 k^2}\bigr)^{1/2}}{32  k_0^4} - \frac{ m^4 \kappa^2 \lambda \log\bigl(1 + \frac{ k_0 k^2}{m^2}\bigr)}{32  k_0^4 k^2} \nonumber \\ && + \frac{  k_1^2 \arctan\Bigl(\frac{ k_0^{1/2} k}{\bigl(4 m^2 -   k_0 k^2\bigr)^{1/2}}\Bigr) k^{3} \bigl(4 m^2 -   k_0 k^2\bigr)^{1/2}}{64  k_0^{9/2}} \Bigg) \nonumber \\ && + \frac{1}{\bar{\phi}}\intp{p} e^{-ip\cdot x} \tilde{\bar{\phi}} \frac{3  k_1 m^4 \kappa^2 \log\Bigl(\frac{1 + \bigl(1 + \frac{4 m^2}{ k_0 p^2}\bigr)^{1/2}}{-1 + \bigl(1 + \frac{4 m^2}{ k_0 p^2}\bigr)^{1/2}}\Bigr) \bigl(1 + \frac{4 m^2}{ k_0 p^2}\bigr)^{1/2}}{32  k_0^4} .
\end{eqnarray}
The logarithmic terms appearing in expressions above are dealt with as follows. In the context of the present problem and the effective theory treatment, we restrict ourselves to the condition $k\ll 10^{-6}M_p$ so that $\frac{k_0 k^2}{m^2}\ll 1$ (more on this later) using the order-of-magnitude estimates of parameters in Eq. (\ref{param}) from the results of \cite{ferreira2018}. Hence, logs involving this fraction can be expanded in a Taylor series. On the other hand, $\sqrt{1 + \frac{m^2}{k_0 k^2}}\approx \sqrt{\frac{m^2}{k_0 k^2}}$. For the $\arctan(\cdots)$ term, we use $\arctan(x)\approx x$ for small $x$. After these expansions, all terms with factors of $k$ will vanish since we assume the derivatives of $\bar{\phi}$ to be zero. Hence, all the integrands of momenta integrals in Eqs. (\ref{efpot1},\ref{b2}) reduce to c-numbers times Fourier transforms of $\bar{\phi}^{n}$. Using these simplifications, the coefficients $A_{2}$ and $B_{2}$ are obtained as,
\begin{eqnarray}
    \label{efpot2}
    A_{2} &=& (\gamma + \log(\pi)) (- \frac{ k_1 m^4}{4  k_0^4} -  \frac{5 m^4 \kappa^2}{32  k_0^2} + \frac{ m^2 \lambda}{8  k_0^3})  + \frac{3 k_1 m^4}{8  k_0^4} + \frac{ m^4 \kappa^2}{4  k_0^2} -  \frac{ m^2 \lambda}{8  k_0^3} \nonumber \\ && + \dfrac{3 m^4 \kappa^2}{32 k_0^2} + (- \frac{ k_1 m^4}{4  k_0^4} -  \frac{5 m^4 \kappa^2}{32  k_0^2} + \frac{ m^2 \lambda}{8  k_0^3}) \log(\frac{m^2}{ k_0 \mu^2}); \nonumber \\
    B_{2} &=& - \frac{9  k_1^2 m^4}{128  k_0^6} -  \frac{5  k_1 m^4 \kappa^2}{16  k_0^4} + \frac{  k_1 m^2 \lambda}{4  k_0^5} + \frac{25 m^2 \kappa^2 \lambda}{192  k_0^3} -  \frac{ \lambda^2}{16  k_0^4} \nonumber \\ &&  -\dfrac{m^2\kappa^2\lambda}{32 k_0^3} + \dfrac{\lambda^2}{32 k_0^4} - \dfrac{m^2 \kappa^2 \lambda }{32 k_0^3} + \dfrac{3 k_1 m^4 \kappa^2}{32 k_0^4} \nonumber \\ && + (\gamma + \log(\pi)) (\frac{3  k_1^2 m^4}{64  k_0^6} + \frac{7  k_1 m^4 \kappa^2}{32  k_0^4} -  \frac{  k_1 m^2 \lambda}{4  k_0^5} -  \frac{13 m^2 \kappa^2 \lambda}{192  k_0^3} + \frac{ \lambda^2}{32  k_0^4}) \nonumber \\ && + (\frac{3  k_1^2 m^4}{64  k_0^6} + \frac{5  k_1 m^4 \kappa^2}{32  k_0^4} -  \frac{  k_1 m^2 \lambda}{8  k_0^5} -  \frac{13 m^2 \kappa^2 \lambda}{192  k_0^3} + \frac{ \lambda^2}{32  k_0^4}) \log(\frac{m^2}{ k_0 \mu^2}) 
\end{eqnarray}
The counterterms for quadratic and quartic terms have a similar form to Eq. (\ref{ren05}), so that the effective potential can be written in terms of renormalized parameters which can be calculated from Eq. (\ref{ren06}) with $A=0$. The effective action takes the form,
\begin{eqnarray}
    \label{efpot3}
    V_{eff} = \dfrac{1}{2}\dfrac{m^2}{k_0} \bar{\phi}^2 + \dfrac{1}{4!}\dfrac{\lambda}{k_0^2}\bar{\phi}^4 + A_{2}\bar{\phi}^2 + B_{2}\bar{\phi}^4.
\end{eqnarray}

\subsection{Estimating the magnitude of corrections}
Making a definitive statement about cosmological implications of quantum corrected potential requires an analysis in the FRW background, which unfortunately is out of scope of the present work. However, we can get an order-of-magnitude estimate of the quantum corrections to the effective potential using the values of parameters $k_0, k_1, m^2, \lambda$ from the results of Ref. \cite{ferreira2018}. 

From the action (\ref{action}), the Einstein equations are given by,
\begin{eqnarray}
    \label{om0}
    3 H^2 &=& \dfrac{\kappa^2}{8}\left(-3\dot{\bar{\phi}}^{2} - 3 \dfrac{k_1}{k_0^2}\bar{\phi}^2\dot{\bar{\phi}}^2 + \dfrac{m^2}{k_0}\bar{\phi}^2 + \dfrac{\lambda}{12 k_0^2}\bar{\phi}^4\right); \nonumber \\
    2\dot{H} + 3 H^2 &=& \dfrac{\kappa^2}{8}\left(-\dot{\bar{\phi}}^{2} - \dfrac{k_1}{k_0^2}\bar{\phi}^2\dot{\bar{\phi}}^2 + \dfrac{m^2}{k_0}\bar{\phi}^2 + \dfrac{\lambda}{12 k_0^2}\bar{\phi}^4\right),
\end{eqnarray}
from which we obtain in the de-Sitter limit ($\dot{H}\sim\dot{\phi}\sim 0$),
\begin{eqnarray}
    \label{om1}
    3H^2 = \dfrac{\kappa^2}{8}\left( \dfrac{m^2}{k_0}\bar{\phi}^2 + \dfrac{\lambda}{12 k_0^2}\bar{\phi}^4 \right). 
\end{eqnarray}
The field equation for $\bar{\phi}$ reads,
\begin{eqnarray}
    \label{om2}
    (a + \frac{k_1}{k_0^2}\bar{\phi}^2)\ddot{\bar{\phi}} + \dfrac{k_1}{k_0^2}\bar{\phi}\dot{\bar{\phi}}^2 + (3aH + \frac{2k_1}{k_0^2}H\bar{\phi}^2)\dot{\bar{\phi}} - \dfrac{m^2}{k_0}\bar{\phi} - \dfrac{\lambda}{6k_0^2}\bar{\phi}^3 = 0.
\end{eqnarray}
Applying the de-Sitter conditions, Eq. (\ref{om2}) yields the de-Sitter value of $\bar{\phi}$, 
\begin{eqnarray}
    \label{om3}
    \bar{\phi}_{0}^2 = - \dfrac{6 k_0 m^2}{\lambda}.
\end{eqnarray}
Using Eq. (\ref{om3}) in (\ref{om1}), we find the de-Sitter value of Hubble parameter $H_{0}$:
\begin{eqnarray}
    \label{om4}
    H_0^2 = - \dfrac{\kappa^2 m^4}{8\lambda}.
\end{eqnarray}
Clearly, the condition for existence of de-Sitter solutions is $\lambda < 0$. Demanding this condition in Eq. (\ref{param}), along with $m^2>0$ and $\bar{\phi}<f$, leads to a constraint on the parameter $\alpha$ of the original theory (\ref{eq02}): $0.5 < \alpha \lesssim 1$. Following the results of \cite{ferreira2018}, we choose $0.5\lesssim\alpha\lesssim 0.6 \sim \mathcal{O}(1)$. Near this value of $\alpha$, $f\sim M_{p}=1/\kappa$ and $\Lambda \sim 10^{16} GeV$. Substituting these in Eqs. (\ref{param}), we find $m^2 \sim \Lambda^{4}/f^{2} \sim 10^{-12}M_{p}^{2}$; $\lambda \sim \Lambda^{4}/f^{4} \sim 10^{-12}$. Similarly, $k_0 \sim 1$ while $k_{1}\sim M_{p}^{-2}$. This also implies that in the low energy limit where momenta $k\ll 10^{13}$ GeV $\ll M_p$, $k_{1}k^2/k_{0}^2 \ll \lambda/k_{0}^2$, i.e. the derivative coupling term is suppressed.

From the above, we can estimate the order of magnitude contributions of terms in $A_{2}$ and $B_{2}$ at $\mathcal{O}(\bar{\phi}^2)$ and $\mathcal{O}(\bar{\phi}^4)$ respectively. We estimate the magnitude of each type of term present at both orders. At quadratic order in background field, we find,
\begin{eqnarray}
    \label{om5}
    \dfrac{\kappa^2 m^4}{k_{0}^2} \sim \dfrac{\lambda m^2}{k_0^3} \sim \dfrac{k_{1}m^4}{k_0^4} \sim 10^{14} GeV^2 .
\end{eqnarray}
Similarly, at quartic order in background field,
\begin{eqnarray}
    \label{om6}
    \dfrac{\kappa^2 m^4 k_1}{k_0^4} \sim \dfrac{\kappa^2 m^2 \lambda}{k_0^3} \sim \dfrac{\lambda^2}{k_0^4} \sim \dfrac{m^4 k_1^2}{k_0^6} \sim \dfrac{k_1 m^2 \lambda}{k_0^5} \sim 10^{-24} .
\end{eqnarray}
Quite an interesting observation here is that the magnitudes of gravitational (terms with a factor of $\kappa^2$) and non-gravitational (terms without $\kappa$) corrections turn out to be exactly the same for both quadratic and quartic order contributions. However, the corresponding quantum corrections are expectedly smaller by an order of $10^{-12}$ compared to $m^2$ and $\lambda$, as can also be checked using the loop counting parameter for de Sitter inflation $H_{0}^{2}/M_{Pl}^{2}$ with $H_0 \sim 10^{13}GeV$ and $M_{Pl}\sim 10^{19}GeV$.

\section{\label{end}Conclusion}
The nonminimal natural inflation model in consideration here is approximately described by a massive scalar field model with quartic self interaction and a derivative coupling in the region where $\phi/f<1$. We study one-loop corrections to this theory, about a Minkowski background, using a covariant effective action approach developed by DeWitt-Vilkovisky. The one-loop divergences and corresponding counterterms have been obtained. Along the way, we also recover several past results, both gauge-invariant non-gauge-invariant, for similar theories. In one such exercise, we obtain the $\phi^4$ coupling correction in a theory with nonminimal coupling of scalar field to gravity, originally considered in Ref. \cite{mackay2010} and thereby extend their result. 

Finite corrections have been taken into account for the calculation of effective potential, where we assume that the background field changes sufficiently slowly so that all derivatives of background field(s) can be ignored. Although cosmologically relevant inferences are not feasible as long as the metric background is Minkowski and not FRW, we can still estimate approximately the magnitudes of quantum corrections. Using the range of parameters applicable to our model, we find that the gravitational and non-gravitational corrections are of same order of magnitudes, while still being expectedly small compared to $m^2$ and $\lambda$. 

This is quite an interesting observation, since one would naively assume that gravitational corrections are $\kappa^2$ suppressed and thus would necessarily be small. There is thus enough motivation to go a step further, and calculate gravitational corrections in the FRW background so that cosmologically relevant inferences can be derived.

\begin{acknowledgments}
 This work was partially funded by DST (Govt. of India), Grant No. SERB/PHY/2017041. 
\end{acknowledgments}

\appendix

\section{\label{a1}Loop Integrals}
Most of the loop integrals are calculated using the well known PV reduction method \cite{bardin1999}. Some integrals, namely (\ref{int4},\ref{int5}) are calculated the general method outlined in Ref. \cite{romao2019}. Finite parts have been calculated for integrals needed for evaluating the effective potential.

Integrals in $\langle\tilde{S}_{2}\rangle$,$\langle\tilde{S}_{4}\rangle$:
\begin{eqnarray}
    \intp{k}\dfrac{k_{\mu}k_{\nu}}{k^2+\frac{m^2}{k_0}} &=& \dfrac{g_{\mu\nu}}{16\pi^2}\Big(\frac{m^4}{8 k_0^2} -  \frac{m^4 \bigl(-1 -  \frac{2}{\epsilon} + \gamma + \log(\pi) + \log(\frac{m^2}{k_0 \mu^2})\bigr)}{4 k_0^2}\Big) \\
    \intp{k}\dfrac{k_{\mu}}{k^2+\frac{m^2}{k_0}} &=& 0 \\
    \intp{k}\dfrac{1}{k^2+\frac{m^2}{k_0}} &=& \dfrac{1}{16\pi^2}\frac{m^2 \bigl(-1 -  \frac{2}{\epsilon} + \gamma + \log(\pi) + \log(\frac{m^2}{k_0 \mu^2})\bigr)}{k_0}
\end{eqnarray}
Integrals in $\langle\tilde{S}_{1}\tilde{S}_{1}\rangle$, $\langle\tilde{S}_{1}\tilde{S}_{3}\rangle$:
\begin{eqnarray}
    \intp{k'}\dfrac{k'^{\mu} k'^{\nu}}{k'^4 ((k'-k)^2+\frac{m^2}{k_0})} &=& \dfrac{1}{16\pi^2} \biggl(\tfrac{1}{4} g^{\mu\nu} \bigl(\frac{2}{\epsilon} -  \gamma -  \log(\pi)\bigr) \nonumber \\ && + \frac{k^{\mu} k^{\nu} \Bigl(\tfrac{1}{2} \bigl(\frac{2}{\epsilon} -  \gamma -  \log(\pi)\bigr) + \tfrac{1}{2} \bigl(- \frac{2}{\epsilon} + \gamma + \log(\pi)\bigr)\Bigr)}{2 k^2}\biggr) \\
    \intp{k'}\dfrac{k'^{\mu} k'^{\nu}}{k'^2 ((k'-k)^2+\frac{m^2}{k_0})} &=& \dfrac{1}{16\pi^2} \tfrac{1}{3} k^{\mu} k^{\nu} \bigl(\frac{2}{\epsilon} -  \gamma -  \log(\pi)\bigr) \nonumber \\ && -  \tfrac{1}{4} g^{\mu\nu} \bigl(\frac{2}{\epsilon} -  \gamma -  \log(\pi)\bigr) \bigl(\frac{m^2}{k_0} + \tfrac{1}{3} k^2\bigr)
\end{eqnarray}

\begin{eqnarray}
    \intp{k'}\dfrac{k'^{\mu}}{k'^2 ((k'-k)^2+\frac{m^2}{k_0})} &=& \dfrac{1}{16\pi^2} \dfrac{k^{\mu}}{2}(\frac{2}{\epsilon}-\gamma - \log(\pi)) \\
    \intp{k'}\dfrac{k'^{2}}{((k'-k)^2+\frac{m^2}{k_0})} &=& \dfrac{1}{16\pi^2} \Bigg[4 \Bigl(\frac{m^4}{8 k_0^2} - \frac{m^4 \bigl(-1 -  \frac{2}{\epsilon} + \gamma + \log(\pi) + \log(\frac{m^2}{k_0 \mu^2})\bigr)}{4 k_0^2}\Bigr) \nonumber \\ && + \frac{m^2 \bigl(-1 -  \frac{2}{\epsilon} + \gamma + \log(\pi) + \log(\frac{m^2}{k_0 \mu^2})\bigr) (k^2)}{k_0}\Bigg] \\
    \intp{k'}\dfrac{1}{k'^2 ((k'-k)^2+\frac{m^2}{k_0})} &=& \dfrac{1}{16\pi^2} \Big(2 + \frac{2}{\epsilon} -  \gamma -  \log(\pi) -  \log(\frac{m^2}{k_0 \mu^2}) \nonumber \\ && -  \log\bigl(1 + \frac{k_0 (k^2)}{m^2}\bigr) \bigl(1 + \frac{m^2}{k_0 (k^2)}\bigr)\Big)\\
    \intp{k'}\dfrac{1}{((k'-k)^2+\frac{m^2}{k_0})} &=& \dfrac{1}{16\pi^2} \frac{m^2 \bigl(-1 -  \frac{2}{\epsilon} + \gamma + \log(\pi) + \log(\frac{m^2}{k_0 \mu^2})\bigr)}{k_0}
\end{eqnarray}
Integrals in $\langle\tilde{S}_{2}\tilde{S}_{2}\rangle$:
\begin{eqnarray}
    \label{int4}
    \intp{k'}\dfrac{k'^{4}}{(k'^2 + \frac{m^2}{k_0}) ((k'-k)^2+\frac{m^2}{k_0})} = \dfrac{1}{16\pi^2} \Bigg(\frac{9 m^4}{16 k_0^2} + \frac{3 m^4}{4 \epsilon k_0^2} -  \frac{3 m^4 \gamma}{8 k_0^2} -  \frac{3 m^4 \log(\pi)}{8 k_0^2} \nonumber \\  -  \frac{3 m^4 \log(\frac{m^2}{k_0 \mu^2})}{8 k_0^2} + \frac{7 m^2 (k^2)}{8 k_0} + \frac{7 m^2 (k^2)}{4 \epsilon k_0} \nonumber \\  - \frac{7 m^2 \gamma (k^2)}{8 k_0} - \frac{7 m^2 \log(\pi) (k^2)}{8 k_0} -  \frac{7 m^2 \log(\frac{m^2}{k_0 \mu^2}) (k^2)}{8 k_0} + \tfrac{1}{8} (k^2)^2 \nonumber \\  + \frac{(k^2)^2}{8 \epsilon} -  \tfrac{1}{16} \gamma (k^2)^2 -  \tfrac{1}{16} \log(\pi) (k^2)^2 -  \tfrac{1}{16} \log(\frac{m^2}{k_0 \mu^2}) (k^2)^2 \nonumber \\ -  \dfrac{\arctan\Bigl(\frac{k_0^{1/2} (k^2)^{1/2}}{\bigl(4 m^2 -  k_0 (k^2)\bigr)^{1/2}}\Bigr) (k^2)^{3/2} \bigl(4 m^2 -  k_0 (k^2)\bigr)^{1/2}}{8 k_0^{1/2}}\Bigg)\\
    \label{int5}
     \intp{k'}\dfrac{k'^{2}k'^{\mu}}{(k'^2 + \frac{m^2}{k_0}) ((k'-k)^2+\frac{m^2}{k_0})} = -\dfrac{1}{16\pi^2} (\frac{2}{\epsilon}-\gamma - \log(\pi))\dfrac{3m^2}{2 k_0}k^{\mu}
\end{eqnarray}

\begin{eqnarray}
\label{int1}
    \intp{k'}\dfrac{k'^{\mu}k'^{\nu}}{(k'^2 + \frac{m^2}{k_0}) ((k'-k)^2+\frac{m^2}{k_0})} = \dfrac{1}{16\pi^2} \nonumber \\ g^{\mu\nu} \biggl(\frac{m^2 \bigl(-1 -  \frac{2}{\epsilon} + \gamma + \log(\pi) + \log(\frac{m^2}{k_0 \mu^2})\bigr)}{6 k_0} + \tfrac{1}{18} \bigl(- \frac{6 m^2}{k_0} -  (k^2)\bigr) \nonumber \\ -  \frac{\bigl(\frac{2}{\epsilon} -  \gamma -  \log(\pi)\bigr) \Bigl(\frac{2 m^4}{k_0^2} + (k^2)^2 - 2 \bigl(\frac{m^4}{k_0^2} -  \frac{2 m^2 (k^2)}{k_0}\bigr)\Bigr)}{12 (k^2)}\biggr) \nonumber \\ + k^{x1} k^{x2} \biggl(\frac{m^2 \bigl(-1 -  \frac{2}{\epsilon} + \gamma + \log(\pi) + \log(\frac{m^2}{k_0 \mu^2})\bigr)}{3 k_0 (k^2)} + \frac{\frac{6 m^2}{k_0} + (k^2)}{18 (k^2)} \nonumber \\ + \frac{\bigl(\frac{2}{\epsilon} -  \gamma -  \log(\pi)\bigr) \Bigl(\frac{2 m^4}{k_0^2} -  \frac{3 m^2 (k^2)}{k_0} + (k^2)^2 - 2 \bigl(\frac{m^4}{k_0^2} -  \frac{2 m^2 (k^2)}{k_0}\bigr)\Bigr)}{3 (k^2)^2}\biggr)\\
    \label{int2}
    \intp{k'}\dfrac{k'^{\mu}}{(k'^2 + \frac{m^2}{k_0}) ((k'-k)^2+\frac{m^2}{k_0})} = -\dfrac{1}{16\pi^2} k^{\mu} \frac{1}{2}\Big(-\frac{2}{\epsilon} + \gamma + \log(\pi)\Big)\\
	\label{int3}     
     \intp{k'}\dfrac{1}{(k'^2 + \frac{m^2}{k_0}) ((k'-k)^2+\frac{m^2}{k_0})} = \dfrac{1}{16\pi^2} \Big(2 + \frac{2}{\epsilon} -  \gamma -  \log(\pi) -  \log(\frac{m^2}{k_0 \mu^2}) \nonumber \\ -  \log\Bigl(\frac{1 + \bigl(1 + \frac{4 m^2}{k_0 (k^2)}\bigr)^{1/2}}{-1 + \bigl(1 + \frac{4 m^2}{k_0 (k^2)}\bigr)^{1/2}}\Bigr) \bigl(1 + \frac{4 m^2}{k_0 (k^2)}\bigr)^{1/2}\Big)
\end{eqnarray}
Integrals of type (\ref{int1},\ref{int2},\ref{int3}) are also present in $\langle\tilde{S}_{1}\tilde{S}_{1}\tilde{S}_{2}\rangle$. The rest of the integrals are, 
\begin{eqnarray}
\intp{k'}\dfrac{k'^{\mu}k'^{\nu}k'^{\rho}}{d_0 d_1 d_2} &=& \dfrac{1}{16 \pi^2} \dfrac{1}{12} \Big(\frac{2}{\epsilon} - \gamma - \log(\pi)\Big) \nonumber \\ && \times(g^{\nu \rho} (2 k^{\mu} + p^{\mu}) + g^{\rho\mu } (2 k^{\nu } + p^{\nu }) + g^{\mu \nu } (2 k^{\rho} + p^{\rho}) \\
\intp{k'}\dfrac{k'^{\mu}k'^{\nu}}{d_0 d_1 d_2} &=& \dfrac{1}{16 \pi^2} g^{\mu\nu} \Big(\frac{2}{\epsilon} - \gamma - \log(\pi)\Big)
\end{eqnarray}


\bibliography{ref,references}

\end{document}